\documentclass{aastex63}

\usepackage{natbib}
\usepackage{hyperref}
\received{December  xx, 2020}
\revised{XXXX}
\accepted{xxx}
\submitjournal{ApJ}
\shorttitle{Chemical analysis of HE2148-2039 and HE 2155$-$2043}
\shortauthors{Purandardas and Goswami}
\graphicspath{{./}{figures/}}

\begin{document}

\title{Chemical  analysis of  two extremely metal-poor stars  HE 2148$-$2039 and HE 2155$-$2043\footnote{Based on data collected at Subaru Telescope, which is operated
by the National Astronomical Observatory of Japan. The data are
retrieved from the JVO portal (\url{http://jvo.nao.ac.jp/portal/v2/})
operated by the NAOJ}}

\correspondingauthor{Aruna Goswami}
\email{aruna@iiap.res.in}
\author[0000-0001-5047-5950]{Meenakshi Purandardas}
\affiliation{Indian Institute of Astrophysics, Koramangala, Bangalore 560034, India;}
\affiliation{Department of physics, Bangalore university, Jnana Bharathi Campus, Karnataka 560056, India}

\author[0000-0002-8841-8641]{Aruna Goswami}
\affiliation{Indian Institute of Astrophysics, Koramangala, Bangalore 560034, India;}

\begin{abstract}
\noindent We present elemental abundance results for HE~2148$-$2039 and HE~2155$-$2043 based on a detailed  high-resolution spectroscopic analysis. The high-resolution SUBARU/HDS spectra used for our analysis have a resolution of R$\sim$60 000. Although limited information based on photometry and low-resolution spectroscopy are available,  we present for the first time an abundance analysis based on high-resolution spectra for both the objects. Our analysis shows that the two objects are extremely metal-poor with [Fe/H]$<$ $-3$. Among the neutron-capture elements, abundances of only Sr and Ba could be determined in our programme stars. For both the objects [Ba/Fe] is found to be $<$ 0. While strontium is under abundant in HE~2148$-$2039 with [Sr/Fe]$\sim$ $-2.02$, Sr is near solar in HE~2155$-$2043. The locations of the programme stars in the absolute carbon abundance, A(C) vs. [Fe/H] diagram show that HE2148$-$2039 is a  CEMP-no Group II object and HE~2155$-$2043 is a CEMP-no Group III object. Observed [Sr/Ba] ratios  are  characteristics of a fast rotating massive star  progenitor for HE2155$-$2043 and  a metal-poor Asymptotic giant branch (AGB) star for HE2148$-$2039. The estimated [Sc/Mn] as well as [C/Cr] ratios in HE2155$-$2043 show that the surface chemical composition of this object is mono-enriched. The surface chemical composition of HE~2148$-$2039 is also found to be mono-enriched based on [Mg/C] ratio.  With respect to their locations in the [C/N] vs. T$_{eff}$ diagram,  HE~2148$-$2039 shows signatures of mixing, and HE2155$-$2043 falls in the unmixed region of [C/N] vs. T$_{eff}$ plot. Kinematic analysis shows that both the objects belong to Galactic halo population.

\end{abstract}

\keywords{stars: abundances--- 
stars: chemically peculiar---stars: Population II}
\section{Introduction} \label{sec:intro}
The Halo of our Galaxy holds a large number of stars that are older than the Milky Way \citep{frebel2018}. Understanding the chemical compositions of Galactic halo stars can give insight into various stellar progenitors and help to better understand the nature of nucleosynthetic pathways at earlier times. First generation stars are thought to be massive stars. They are expected to synthesize elements  up to iron peak including Sr or Ba by weak slow neutron-capture (-s) process \citep{Frischknecht.2012A&A...538L...2F}. About half of the heavy elements were produced only after the low- and intermediate- mass stars were formed and evolved to their asymptotic giant branch (AGB) phase 
 \citep{lugaro.2003ApJ...586.1305L,herwig.2005ARA&A..43..435H,karakas.2014PASA...31...30K}. Core collapse supernovae, neutron star mergers etc. are the other known sources of heavy elements, the rapid neutron-capture (r-) process elements \citep{arcones.2013JPhG...40a3201A, rosswog.2014MNRAS.439..744R, abbott.2017ApJ...848L..13A, drout.2017Sci...358.1570D, shappee.2017Sci...358.1574S}.

\par Metal-poor ([Fe/H] $<$ $-1.0$) stars with enhanced carbon ([C/Fe] ${\geq}$ 0.7, \cite{aoki.2007ApJ...655..492A}) are  known as carbon-enhanced metal-poor (CEMP) stars. Various sky survey programmes such as HES/HK surveys \citep{christlieb.2001A&A...375..366C, wisotzki.2000A&A...358...77W,beers.1985AJ.....90.2089B,beers.1992AJ....103.1987B,beers.1999aASPC..165..202B,beers.1999bAp&SS.265..547B} dedicated to explore the metal-poor stars revealed that the fraction of carbon-enhanced metal-poor stars increases with decrease in metallicty \citep{cohen.2005ApJ...633L.109C,frebel.2006aApJ...638L..17F,carollo.2012ApJ...744..195C,lee.2013AJ....146..132L,placco.2014aApJ...790...34P,beers2017ApJ...835...81B}. CEMP stars show different levels of enhancement of neutron-capture elements. Based on the enhancement of neutron-capture elements, CEMP stars are classified into different groups: CEMP-s, CEMP-r, CEMP-r/s and CEMP-no \citep{beers&christlieb2005ARA&A..43..531B}. 
A detailed study and a comprehensive analysis of the classification criteria put forward by different authors for CEMP stars, as well as production scenarios can be found in \cite{partha.2021arXiv210109518G}.   CEMP-s stars exhibit the enhancement of s-process elements ([Ba/Fe] $>$ 1.0, and [Ba/Eu] $>$ 0.5). CEMP-r/s stars show over abundances of both s- and r- process elements (0.0 $<$ [Ba/Eu] $<$ 0.5). The widely accepted scenario to explain the observed enhancement of heavy elements in these groups of CEMP stars involves a binary mass transfer process  \citep{herwig.2005ARA&A..43..435H, Bisterzo.2012MNRAS.422..849B}. According to this scenario, these stars accreted the heavy elements synthesised by the companion during its AGB phase. Radial velocity variations exhibited by CEMP-s and CEMP-r/s stars also support the binary mass transfer scenario \citep{McClure&Woodsworth1990ApJ...352..709M,hansen.2016bA&A...588A..37H}. CEMP-r stars are the most rarely observed objects and exhibit enhancement of r-process elements ([Eu/Fe] $>$ 1.0). 
CS 22892-052 is the CEMP-r star first reported by \cite{McWilliam.1995AJ....109.2757M} and \cite{sneden.2003ApJ...591..936S} which is found to be unlikely in a binary system \citep{hansen.2011ApJ...743L...1H}. \cite{hansen.2015ApJ...807..173H} compared the elemental abundance patterns estimated in two CEMP-r stars HE 0010$-$3422 and HE 0448$-$4806, in their sample with the yields from  metal-poor AGB stars with different masses.  Massive AGB models could reproduce only some of the observed abundances for the neutron capture elements in HE 0010$-$3422 which requires radial velocity monitoring to confirm its binarity. For HE 0448$-$4806,  none of the AGB models could give a good fit for the observed abundances.  \cite{hansen.2015ApJ...807..173H} suggest that the observed abundance anomalies in CEMP-r stars are not the result of mass transfer from a binary companion but may be inherited from the interstellar medium from which the star was formed.
\par Unlike the other groups of CEMP stars, CEMP-no stars do not show any enhancement of neutron-capture elements. CEMP-no stars constitute ${\sim}$ 20\% of very metal-poor ([Fe/H] $<$ $-2.0$) stars, ${\sim}$ 40\% of extremely metal-poor ([Fe/H] $<$ $-3.0$) stars,  and ${\sim}$ 80\% of ultra metal-poor ([Fe/H] $<$ $-4.0$) stars and the fraction increases with lower metallicities \citep{banerjee.2018MNRAS.480.4963B}. CEMP-no stars mostly occur as single stars \citep{norris.2013ApJ...762...28N,starkenburg.2014MNRAS.441.1217S,hansen.2016aA&A...586A.160H}. Only a very few of the CEMP-no stars are known to be confirmed binaries. \cite{hansen.2016aA&A...586A.160H} reported that 17\% of CEMP-no stars in their sample of 24 stars are binaries. \cite{arentsen.2019A&A...621A.108A} suggest that the CEMP-no star may be in a binary system and the companion is an extremely metal-poor star and once passed through the AGB phase but that has not produced any significant amount of s-process elements. Another scenario proposed to explain the origin of CEMP-no stars is that they are formed from an interstellar medium polluted by Spin stars \citep{meynet.2010A&A...521A..30M,chiappini.2013AN....334..595C} or Faint supernovae that underwent mixing and fall back \citep{umeda.2005ApJ...619..427U,nomoto.2013ARA&A..51..457N,tominaga.2014ApJ...785...98T} or metal-free massive stars \citep{heger.woosely.2010ApJ...724..341H}. \cite{hansen.2016aA&A...586A.160H} suggest that all these sources may contribute towards the observed abundance patterns of CEMP-no stars. CEMP-no stars are expected to be the direct descendants of the first generation stars \citep{ito.2013ApJ...773...33I,spite.2013,placco.2014aApJ...790...34P,hansen.2016A&A...588A..37H}. Many studies show that the surface composition of CEMP-no stars match well with that expected for population III stars \citep{christlieb.2004A&A...428.1027C,meynet.2006A&A...447..623M,frebel.2008ApJ...684..588F,nomoto.2013ARA&A..51..457N,keller.2014Natur.506..463K,bonifacio.2015A&A...579A..28B,yoon.2016ApJ...833...20Y,placco.2016ApJ...833...21P,choplin.2017A&A...607L...3C,Ezzeddine.2019ApJ...876...97E}.

\par Since the CEMP-no stars are the bonafide second generation stars \citep{placco.2015ApJ...809..136P,hansen.2016aA&A...586A.160H,yoon.2016ApJ...833...20Y}, the abundance analysis of these stars can give important clues about the early Galactic nucleosynthesis. In this work,  we have presented a detailed abundance analysis of two CEMP-no stars, HE 2148$-$2039 and HE 2155$-$2043 using high-resolution spectra. The paper is organized as follows: Section \ref{sec:pre-studies} presents a brief summary of previous works found in literature on the two programme stars. The source of the spectra is presented in section \ref{sec:source-spectra}. Determination of stellar atmospheric parameters, mass and age of the programme stars are discussed in section \ref{sec:stellar-parameters}. Abundance analysis is discussed in section \ref{sec:abundance-analysis}. Section \ref{sec:discussion} presents a discussion on our results. Kinematic analysis is discussed in section \ref{sec:kinematic-analysis}. Conclusions are drawn in Section \ref{sec:conclusion}.
\section{Previous studies: A brief summary} \label{sec:pre-studies}
U, B and V magnitudes for the two objects HE~2148$-$2039 and HE~2155$-$2043 are reported in \cite{beers.2007ApJS..168..128B}. \cite{frebel.2006bApJ...652.1585F} reported a metallicity [Fe/H] $\sim$ $-3.55$ and a carbon abundance [C/Fe] $\sim$ 1.56 for the object HE~2155$-$2043 based on medium resolution ($\sim$ 2{\rm \AA}) spectroscopy. For the object HE~2155$-$2043, \cite{hansen.2016bA&A...588A..37H} reported atmospheric parameters T$_{eff}$ = 5200 K, log g = 2.4, and  $\zeta$ = 1.5 kms$^{-1}$ with a metallicity [Fe/H] $\sim$ $-3.0$. They have determined the abundances of strontium, nitrogen and carbon for this object based on the analysis of medium resolution (R$\sim$ 7450) spectra obtained with X-shooter spectrograph. They obtained [Sr/Fe] $\sim$ 0.20, [N/Fe] $\sim$ 0.0 and [C/Fe] $\sim$ 0.70. They could not determine the abundance of Ba as its absorption features were too weak in the spectrum of this object. They classified the object HE~2155$-$2043 as CEMP-no star based on their analysis. The most recent study available for the object HE~2155$-$2043 is by \cite{beers2017ApJ...835...81B}. They have reported the atmospheric parameters T$_{eff}$ = 5016 K, log g = 2.87, [Fe/H] $\sim$ $-3.27$ and carbon abundance of [C/Fe] $\sim$ 0.81 based on an analysis of medium resolution (R $\sim$ 2000) spectra using non-SEGUE Stellar Parameter Pipeline (n-SSPP). They have also determined a radial velocity of $\sim$ $-92$ kms$^{-1}$ for this object. Although some aspects of our programme stars have been addressed in previous studies, results based on high-resolution spectroscopic analysis are missing. We present for the first time an abundance analysis for both the objects based on high-resolution spectra.
\section{Source of spectra} \label{sec:source-spectra}
\begin{figure}
\centering
\includegraphics[width=11cm,height=9cm]{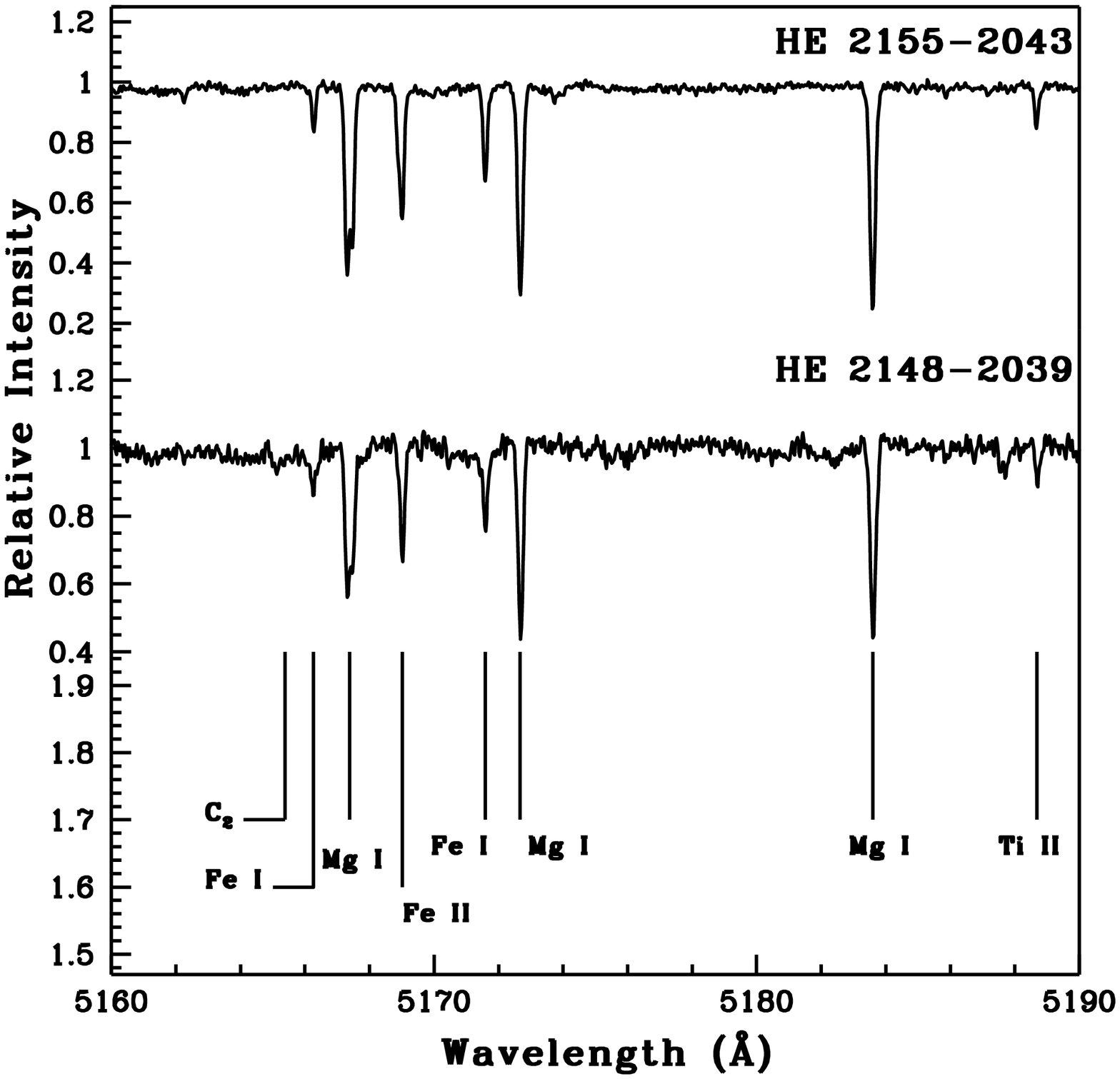}
\includegraphics[width=11cm,height=9cm]{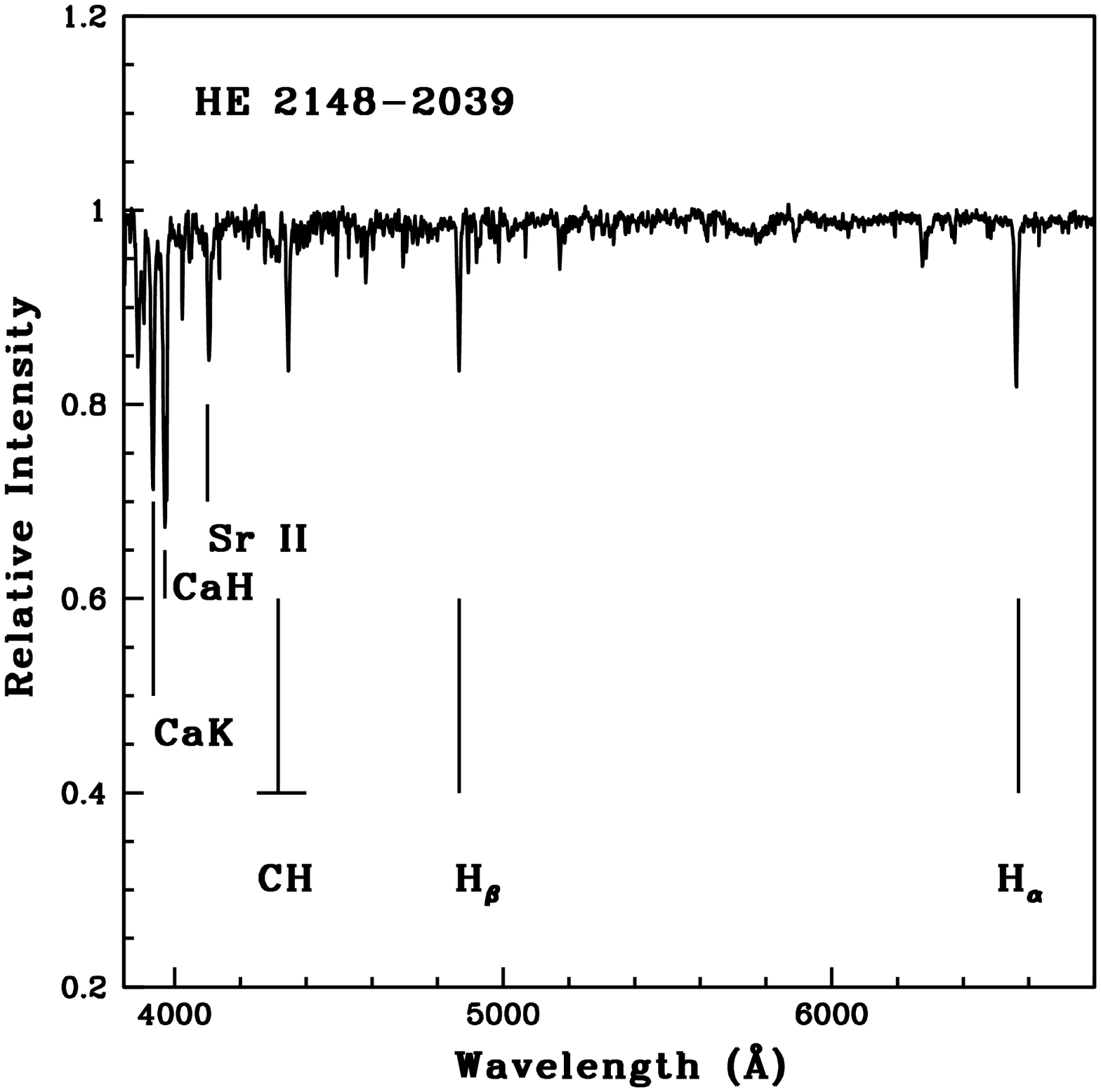}
\caption{ Upper panel: High resolution  spectra of the programme stars  in the  
wavelength region 5160 to 5190 {\bf  {\rm \AA}}.\\
Lower panel:  A low resolution spectrum  of  HE 2148$-$2039 in the  
wavelength region 3840 to 6800 {\bf  {\rm \AA}} acquired with HCT/HFOSC.  }
\label{fig1}
\end{figure}

Programme stars are selected from the list of metal-poor candidates of the Hamburg/ESO survey \citep{christlieb.2003RvMA...16..191C}. We have used the high resolution spectra of the two programme stars obtained using the High Dispersion Spectrograph (HDS) of the 8.2 m Subaru Telescope at a resolution $\sim$ 60 000. The wavelength calibrated spectra were retrieved from the JVO portal (\url{http://jvo.nao.ac.jp/portal/v2/}) operated by National Astronomical Observatory of Japan where the  data are publicly available. The wavelength coverage of the spectra spans from 4100 to 6850{\rm \AA}, with a small gap between 5440 and 5520{\rm \AA}. This gap occurs due to the physical separation between the two CCDs. The detectors are two EEV CCDs with 2048$\times$4096 pixels with two by two on-chip binning. Examples of sample spectra are shown in Figure \ref{fig1} (Upper panel). The basic data for the programme stars are listed in Table \ref{table1}. A low resolution spectrum of the programme star HE 2148$-$2039 obtained using the Himalayan Faint Object Spectrograph Camera (HFOSC) attached to 2-m Himalayan Chandra Telescope (HCT) at the Indian Astronomical Observatory (IAO), Hanle on 20 October 2020 is also shown in Figure \ref{fig1} (Lower panel), that covers a wavelength range from 3840 to 6800{\rm \AA}. HFOSC is an optical imager cum spectrograph used for low- and medium-resolution grism spectroscopy. We have obtained spectra in grism 7 and 8 that cover the wavelength regions from 3800 - 6840 {\rm \AA} with a resolution of R $\sim$ 1330 and 5800 - 9000 {\rm \AA} with a resolution of R $\sim$ 2190 respectively. 

{\footnotesize
\begin{table*}
\caption{\bf Basic data for the programme stars}
\resizebox{\textwidth}{!}{\begin{tabular}{lcccccccccc}
\hline
Star      &RA$(2000)$       &Dec.$(2000)$    &B   &V   &J        &H        &K  &Exposure   &Date of obs. & Source  \\
          &                 &                &        &        &         &         &   & (seconds) &  & of spectrum\\
\hline
HE~2148$-$2039 & 21 51 06.85 & $-20$ 25 48.26 & 14.67$\pm$0.03$^{a}$ & 14.03$\pm$0.02$^{a}$ & 12.48$\pm$0.02 & 12.03$\pm$0.02 & 11.96$\pm$0.02 & 1800 & 08-12-2003 & Subaru/HDS  \\
             &             &                &   13.99$\pm$0.03$^{b}$   &  14.72$\pm$0.04$^{b}$   &       &       &       & 2700 & 20-10-2020 & HCT/HFOSC\\                                         
HE~2155$-$2043 & 21 58 42.27 & $-20$ 58 42.27 & 14.10$^{a}$ &  13.15$\pm$0.02$^{a}$ & 11.57$\pm$0.02 & 11.09$\pm$0.02 & 11.02$\pm$0.02 & 1800 & 27-06-2004 & Subaru/HDS\\ 
             &             &         &      13.19$\pm$0.03$^{b}$  &   13.96$\pm$0.01$^{b}$  &      &       &       &       &  &  \\ 
\hline

\end{tabular}}
\label{table1}

$^{a}$ SIMBAD, $^{b}$ APASS DR10 catalog
\end{table*}
}

\section{Spectroscopic stellar parameters}\label{sec:stellar-parameters}
Radial velocities of the programme stars are determined by measuring the shift in the wavelength for a large number of unblended and clean lines in the spectra. Estimated radial velocities are presented in Table \ref{table2}. 
\par
\begin{figure}
\centering
\includegraphics[width=11cm,height=10cm]{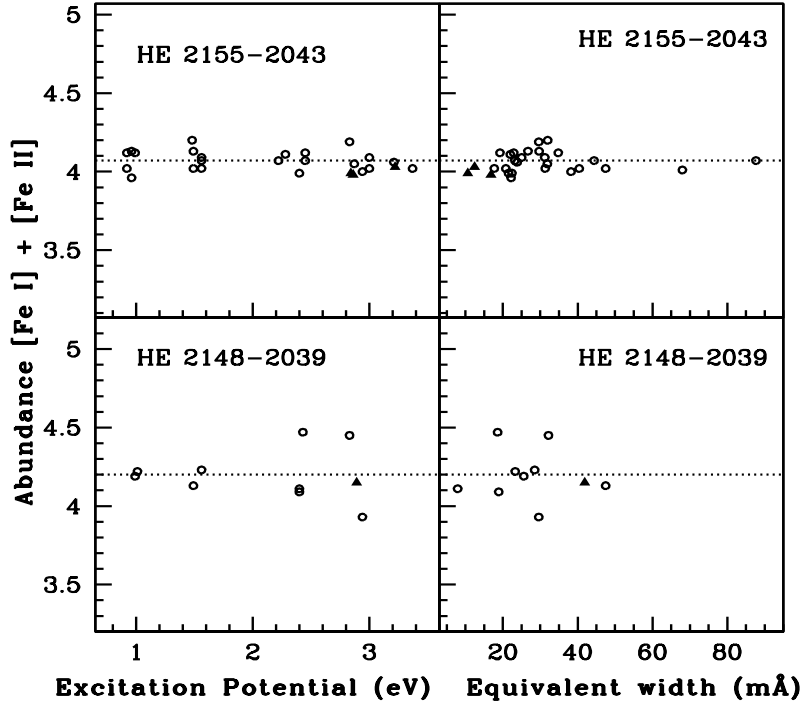}
\caption{ The iron abundances of programme stars as a function of excitation potential (left panels) and equivalent width (right panels). In all the panels, the open circles indicate Fe I lines and solid triangles represent Fe II lines.}
\label{fig2} 
\end{figure}

 Stellar parameters are determined following the detailed procedures described in our earlier paper \cite{purandardas.2019aMNRAS.486.3266P}. Here we mention briefly the main points. Atmospheric parameters of the programme stars are determined from the measured equivalent widths of a number of clean and unblended Fe I and Fe II lines. The number of (Fe I and Fe II) lines measured on the stars' spectra are limitted and the lines are found to be very weak; (9, 1) and (25, 3) respectively for HE 2148$-$2039 and HE 2155$-$2043. Only those lines with excitation potential in the range 0.0 - 5.0 eV are considered for the analysis. We have made use of the recent version of MOOG (Sneden 1973, updated version 2013) for the analysis assuming the local thermodynamic equilibrium (LTE). Model atmospheres are selected from Kurucz grid of model atmospheres with no convective overshooting (\url{http://cfaku5.cfa.hardvard.edu/}).  Solar abundances are adopted from \cite{asplund.2009ARA&A..47..481A}. The effective temperature of a programme star is taken to be that value for which the trend between the abundances obtained from Fe I lines and the excitation potential gives zero slope. A trend between the abundances derived from Fe I lines and the reduced equivalent width with a zero slope defines the microturbulent velocity of the star. Under this temperature and microturbulent velocity, the surface gravity log g, is determined for which the abundances obtained from the Fe I and Fe II lines are nearly the same. The derived atmospheric parameters and radial velocities of the programme stars are listed in Table \ref{table2}. The derived T$_{eff}$ estimates when  compared with  the estimates from   Gaia DR2 are found to be  lower by 314 K in the case of HE~2148$-$2039 and by 382 K in the case of HE~2155$-$2043. This discrepancy may be due to the fact that the  RMS error in T$_{eff}$  provided by Gaia DR2  is found to increase with decreasing  metallicity of  the stars with smallest error around solar metallicity. The Gaia T$_{eff}$ estimates are  	good in  the metallicity range $-$2 $\leq$ [Fe/H] $\leq$ 0.5  as their training sample of stars are in this metallicity range. Outside this range of metallicity, the T$_{eff}$ estimates are found to be more biased and overestimated \citep{gaia.2018A&A...616A...1G}.  The plot showing the estimated abundances from Fe I and Fe II lines as a functions of excitation potential and equivalent widths are shown in Figure \ref{fig2}. The equivalent widths of the Fe I and Fe II lines used for deriving the atmospheric parameters are presented in Table \ref{table3}.

{\footnotesize
\begin{table*}
\caption{\bf Derived atmospheric parameters and radial velocities of  the programme stars.}
\begin{tabular}{lcccccccc}
\hline
Star         &T$_{eff}$& log g &$\zeta$      & [Fe I/H]        &[Fe II/H]        & V$_{r}$       & V$_{r}$&    Ref\\
             &    (K)  & cgs  &(km s$^{-1}$) &                 &                 & (km s$^{-1}$) & (km s$^{-1}$)&    \\
             &($\pm$100)& ($\pm$0.2)     & ($\pm$0.2)&       &   &                 & (SIMBAD)       &             \\
\hline
HE~2148$-$2039 & 5000    & 2.0  & 2.80         & $-3.30$$\pm$0.14& $-3.35$         & 14.95 $\pm$ 0.14 & - &  1\\
               & 5314  &      &               &                 &             &                   &   &  2 \\
HE~2155$-$2043 & 4800    & 0.4  & 1.21         & $-3.43$$\pm$0.06& $-3.49$$\pm$0.04& $-68.92$ $\pm$ 0.53& $-92$ & 1 \\
               & 5182    &      &              &                &                  &                    &      &  2   \\
\hline
\end{tabular}

Ref: 1: This work;  2: Gaia DR2
\label{table2}
\end{table*}
}

{\footnotesize
\begin{table*}
\caption{\bf Equivalent widths (in m\r{A}) of Fe lines used for deriving atmospheric parameters.}
\resizebox{\textwidth}{!}{\begin{tabular}{cccccc}
\hline                       
Wavelength(\r{A}) & Element & $E_{low}$(eV) & log gf & HE~2148$-$2039 & 
HE~2155$-$2043 \\ 
                  &         &               &        &              &             \\
\hline 
4143.869	&Fe I   &	1.557	&	-0.45	&	-	&	87.7(4.07)	\\
4147.670    &		&	1.484	&	-2.10	&	-	&	31.9(4.20)		\\
4187.038	&		&	2.449	&	-0.55	&	-	&	44.4(4.07)		\\
4337.046	&		&	1.557	&	-1.69	&	-	&	40.4(4.02)		\\
4427.309	&		&	0.051	&	-3.04	&	-	&	67.9(4.01)		\\
4447.718	&		&	2.222	&	-1.34	&	-	&	23.2(4.07)		\\
4466.551	&		&	2.832	&	-0.59	&	-	&	29.6(4.19)	  	\\
4531.147	&		&	1.485	&	-2.15	&	-	&	29.7(4.13)		\\
4871.317	&		&	2.868	&	-0.41	&	-	&	31.9(4.05)	  	\\
4994.129	&		&	0.915	&	-3.08	&	-	&	22.9(4.12)	  \\
5006.117	&		&	2.832	&	-0.77	&	32.2(4.45)	&	-		\\
5049.819	&		&	2.279	&	-1.42	&	-	&	22.0(4.11)	  	\\
5051.634	&		&	0.915	&	-2.79	&	-	&	31.3(4.02)	\\
5083.338	&		&	0.958	&	-2.96	&	-	&	26.7(4.13)		\\
5166.281	&		&	0.000	&	-4.19	&	-	&	21.5(3.99)  \\
5171.595	&		&	1.485	&	-1.79	&	47.5(4.13)	&	47.5(4.02)\\
5192.343	&		&	2.998	&	-0.52	&	-	&	20.8(4.02)	      \\
5194.941	&		&	1.557	&	-2.09	&	28.5(4.23)	&	31.2(4.09)\\
5232.939	&		&	2.940	&	-0.19	&	29.6(3.93)	&	38.2(4.00)\\
5266.555	&		&	2.998	&	-0.49  &	-	&	24.9(4.09)	\\
5324.178	&		&	3.211	&	-0.24	&	-	&	23.8(4.06)		\\
5497.516	&		&	1.011	&	-2.85	&	23.3(4.22)	&	-		\\
5501.464	&		&	0.958	&	-2.95	&	-	&	22.2(3.96)	  		\\
5506.778	&		&	0.990	&	-2.79	&	25.6(4.19)&	34.7(4.12)		\\
5586.756	&		&	3.368	&	-0.21	&	-	&	17.7(4.02)		\\
6136.615	&		&	2.453	&	-1.40	&	-	&	19.2(4.12)	\\
6252.554	&		&	2.404	&	-1.69	&	15.0(4.11)	&	-			\\
6393.602	&		&	2.433	&	-1.62	&	18.6(4.47)	&	-		\\
6494.980 	&		&	2.404	&	-1.27	&	18.9(4.09)	&	22.4(3.99)	\\
4508.288	&Fe II	&	2.856	&	-2.21	&	-	&	16.8(3.98)	\\
4515.339	&		&	2.844	&	-2.48	&	-	&	10.5(3.99)		\\
4923.927	&		&	2.891	&	-1.32	&	41.9(4.15)	&	-	\\
5234.625	&		&	3.221	&	-2.05	&	-	&	12.4(4.03)		\\

\hline
\end{tabular}}
		
The numbers in the  parenthesis in columns 5-6 give the derived abundances from the respective line. \\
log gf values are taken from  kurucz atomic line list (\url{https://www.cfa.harvard.edu}).\\
\label{table3}
\end{table*}
}
{\footnotesize
\begin{table*}
\caption{\bf Estimates of  log\,{g} using parallax method }
\begin{tabular}{lccccccc}
\hline                       
 Star name    & Parallax        & $M_{bol}$    & log(L/L$_{\odot}$) & Mass(M$_{\odot}$) & log g & log g (Spectroscopic)  \\
              & (mas)           &              &                    &                   & (cgs)  & (cgs)  \\
\hline
HE~2148$-$2039 & 0.12$\pm$0.03 & $-0.77$$\pm$1.00 & 2.21$\pm$0.21 & 0.95$\pm$0.55 & 1.98$\pm$0.13 & 2.0$\pm$0.2 \\
HE~2155$-$2043 & 0.13$\pm$0.05 & $-1.63$$\pm$0.52 & 2.55$\pm$0.40 & 1.00$\pm$0.30 & 1.54$\pm$0.28 & 0.4$\pm$0.2 \\
\hline
\end{tabular}
\end{table*}
\label{table4}
}

\subsection{Mass and age}
\begin{figure}
\centering
\includegraphics[width=12cm,height=11cm]{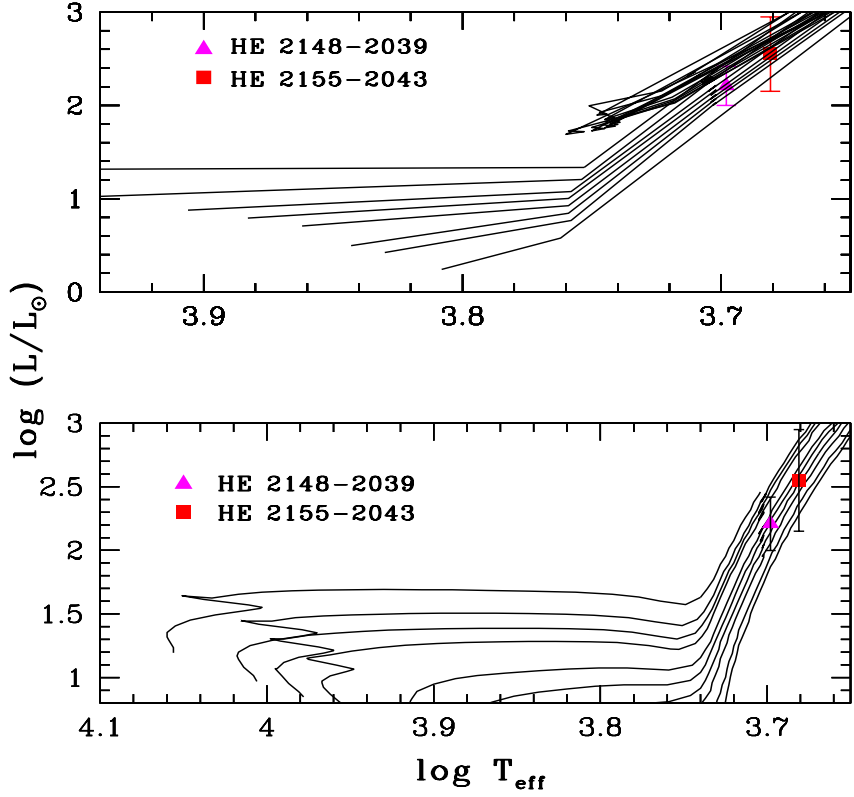}
\caption{The locations of HE 2148$-$2039 and HE 2155$-$2043 in the H-R diagram are shown. The isochrone tracks for log(age) 10.20, 10.0, 9.9, 9.8, 9.70, 9.60, 9.40 and 9.25 are shown from bottom to top in the upper panel. The evolutionary tracks for 0.6, 0.7, 0.8, 0.9, 1.0, 1.1, 1.3, 1.4, 1.5 and 1.7 M$_{\odot}$ are shown from bottom to top in the bottom panel.}
\label{fig3}
\end{figure}

Mass and age of the programme stars are estimated from their positions in the H-R diagram, log(L/L$_{\odot}$) vs. T$_{eff}$.
Luminosities of the stars are determined using the relation, 

 \[  log(L/L_{\odot}) = (M_{\odot}-M_{bol})/2.5\] 

where M$_{\odot}$ is the bolometric magnitude for the sun, and
\[M_{bol} = Mv+BC-Av\]
Mv is calculated using the equation, 
\[Mv = V-(5log(d))+5\]
The visual magnitude (V) of the star is taken from SIMBAD and the parallax values are adopted from \textit{Gaia} (Gaia collaboration et al. 2016, 2018b, \url{https://gea.esac.esa.int/archive/}). Bolometric corrections are calculated using the empirical calibrations of \cite{alonso.1999A&AS..140..261A}. Interstellar extinction required for the estimation of bolometric magnitude is calculated using the formula as given by \cite{chen.1998A&A...336..137C}. The evolutionary tracks and isochrones corresponding to Z = 0.0008 are taken from the \cite{girardi.2000A&AS..141..371G} database to estimate the mass and age of the stars. Locations of the programme stars in the H-R diagram (Figure \ref{fig3}) show that they are in the ascending stage of the giant branch with near solar masses. The estimated ages of the objects HE~2148$-$2039 and HE~2155$-$2043 are 5.6 and 5.0 Gyrs respectively as derived from the parallax method. However, we note that,  the errors in the age estimates are as high as  7Gyr for both the objects which may be due to the high errors in the parallax estimates that are 
${\sim}$ 38\% and 25\% for HE~2155$-$2043 and HE~2148$-$2039 respectively. The parallax error propagates to  the luminosity estimates that could cause uncertainties in accurate age determination from the isochrones in the H-R diagram. \\

\par The surface gravity, log g, is calculated using the relation, \\
\[log (g/g_{\odot}) = log (M/M_{\odot}) + 4log (T_{eff}/T_{eff\odot}) + 0.4(M_{bol} - M_{bol\odot})\] 
Where log g$_{\odot}$ = 4.44, T$_{eff\odot}$ = 5770K, and M$_{bol\odot}$ = 4.75 mag \citep{yang.2016RAA....16...19Y}.\\ The estimated mass and surface gravity  of the programme stars are tabulated in Table \ref{table4}. 

\section{Abundance analysis}\label{sec:abundance-analysis}
Abundance analyses of the programme stars are performed using the measured equivalent widths as well as using the spectral synthesis  calculation of clean and symmetric lines due to various elements. Lines are identified by overplotting the Arcturus spectrum on the individual spectrum of the programme stars. A master linelist is prepared with the measured equivalent widths and line information such as excitation potential and log gf values of the spectral lines taken from the Kurucz database (\url{https://www.cfa.harvard.edu}). Abundances of light elements, C and N, odd-Z element Na, $\alpha$ elements Mg, Ca, Sc and Ti and Fe-peak elements Cr, Mn, Co and Ni are determined whenever possible. Among the neutron-capture elements, we could estimate the abundances of only Sr and Ba. Lines due to other heavy elements are either absent or too weak to measure. The lines used to derive the abundances of various elements are presented in Table \ref{table5}.

\begin{table*}
\caption{\bf Lines used for deriving elemental abundances}
\resizebox{\textwidth}{!}{\begin{tabular}{cccccc}
\hline                       
Wavelength(\r{A}) & Element & $E_{low}$(eV) & log gf & HE~2148$-$2039 & 
HE~2155$-$2043  \\ 
                  &         &               &        &              &               \\
\hline 
5889.951	&	Na I	&	0.000   &	0.10 	&	95.4(2.65)	&	131.1(4.06)	\\
5895.920	&		    &	0.000   &	-0.20	&	73.1(2.69)	&	114.0(3.94)	  	\\
5172.684	&	Mg I	&	2.711	&	-0.40   &	118.7(4.08)	&	141.8(5.59)	\\
5183.600	  &		    &	2.720	&	-0.18	&	123.2(3.93)	&		-         \\
4318.652	&	Ca I	&	1.899	&	-0.21   &	-	        &	19.3(3.48)		\\
6122.217	&		    &	1.885	&	-0.41   &	-	        &	20.0(3.47)	  	\\
6162.173	&		    &	1.899	&	0.10	&	26.0(3.11)	&	28.4(3.20)		\\
6439.075	&		    &	2.525	&	0.47	&	16.8(3.19)	&	18.6(3.26)		\\
4320.732	&	Sc II	&	0.605	&	-0.26	&	29.8(-0.25)	&	43.6(-0.28)\\
4374.457	&		    &	0.618	&	-0.44	&	-	        &	31.7(-0.40)	\\
4415.557	&		    &	0.595	&	-0.64	&	-	        &	26.2(-0.39)	\\
4417.719	&	Ti II	&	1.164	&	-1.43	&	22.0(1.77)	&	40.8(1.85)			\\
4443.794	&		    &	1.080	&	-0.70	&	-	        &	65.1(1.74)		\\
4468.507	&		    &	1.130	&	-0.60	&	-	        &	60.4(1.54)		\\
4563.761	&		    &	1.221	&	-0.96	&	50.4(1.83)	&	46.1(1.54)		\\
5226.540  	&	        &	1.565	&	-1.30	&	16.6(1.88)	&		-       		\\
4289.717	&	Cr I	&	0.000   &	-0.36   &	-	        &	63.1(2.58)	\\
4030.753	&	Mn I	&	0.000   &	-0.47	&	-	        &	65.7(2.27)	\\
4121.311	&	Co I	&	0.922	&	-0.32	&	-	        &	38.9(2.18)		\\
5476.900	  &	Ni I	&	1.826	&	-0.89	&	-	        &	30.6(3.13)\\
4077.709	&	Sr II	&	0.000   &	0.17	&	-	        &	95.1(-0.30)		\\
4215.519	&		    &	0.000   &	-0.14  &	-	        &	88.4(-0.26)		\\
4934.076	&	Ba II	&	0.000   &	-0.15	&	41.5(-1.67)	&	14.9(-2.72)	\\

\hline
\end{tabular}}
The numbers in the  parenthesis in columns 5-6 give the derived abundances from the respective line. log gf values are taken from  kurucz atomic line list (\url{https://www.cfa.harvard.edu})
\label{table5}
\end{table*}

\subsection{Carbon and Nitrogen}
\begin{figure}
\centering
\includegraphics[width=12cm,height=11cm]{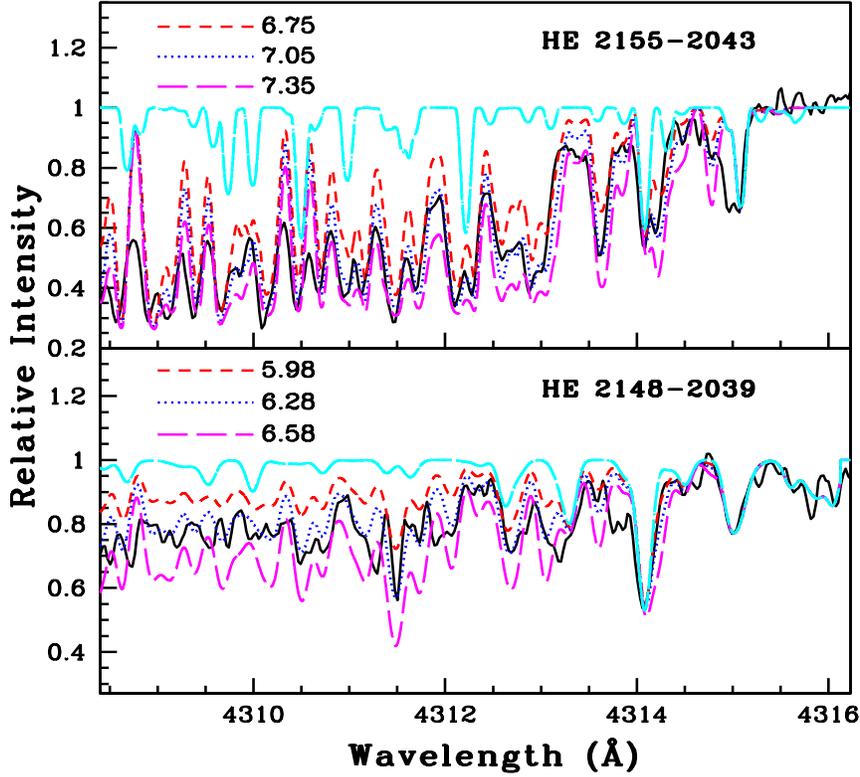}
\caption{Synthesis of CH band around 4315 {\rm \AA}. Dotted line represents synthesized spectra and the solid line indicates the 
observed spectra. Short dashed line represents the synthetic spectra corresponding to $\Delta$ [C/Fe] = -0.3 and long dashed line  corresponds to $\Delta$[C/Fe] = +0.3. Solid cyan line corresponds to a spectral synthesis fit obtained  without CH.}
\label{fig4}
\end{figure}

The abundance of carbon is determined using the spectrum synthesis calculation of CH band at 4315 {\rm \AA}. In both the programme stars, CH bands are clearly detectable. Hence we have used the carbon abundance obtained from the synthesis of CH band throughout our analysis. The spectrum synthesis calculation of CH band gives [C/Fe] $\sim$ 1.15 and 2.05 respectively for HE~2148$-$2043 and HE~2155$-$2043. As the CH band region of  HE~2148$-$2043 is found to be very noisy, the carbon abundance obtained for this object is taken as the upper limit. The best spectrum synthesis fit that we could obtain for the CH band for HE~2148$-$2043 is shown in Figure \ref{fig4}. The C$_{2}$ bands at 5165 and 5635 {\rm \AA} are weak and marginally detectable.  

\begin{figure}
\centering
\includegraphics[width=12cm,height=11cm]{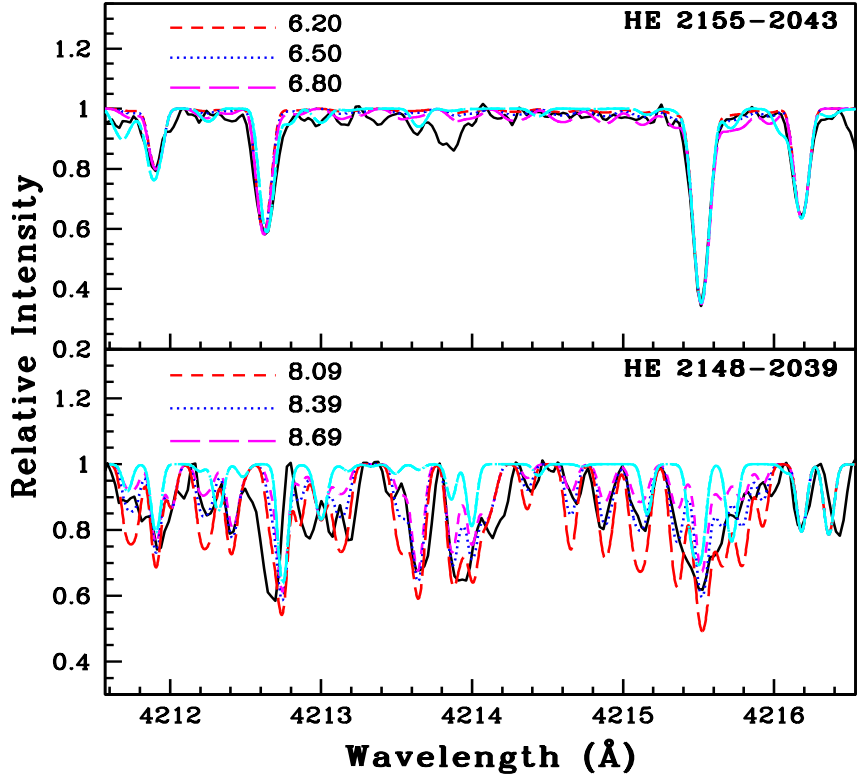}
\caption{ Synthesis of CN band around 4215 {\rm \AA}. Dotted line represents synthesized spectra and the solid line indicates the observed spectra. Short dashed line represents the synthetic spectra corresponding to $\Delta$[N/Fe] = -0.3 and long dashed line  corresponds to $\Delta$[N/Fe] = +0.3. Solid cyan line corresponds to a spectral synthesis fit obtained without CN.}
\label{fig5}
\end{figure}

Abundance of nitrogen is estimated using the spectrum synthesis calculation of CN band at 4215 {\rm \AA}. Nitrogen is found to be enhanced in both the stars with [N/Fe] $\sim$ 3.86 and 2.10 in HE~2148$-$2039 and HE~2155$-$2043 respectively. As the region around 4215 {\rm \AA} is noisy in HE~2148$-$2039, the nitrogen abundance obtained for this star is taken as the upper limit. The best spectrum synthesis fit that we could obtain for the CN band for this object is shown in Figure \ref{fig5}. The molecular linelists for carbon and nitrogen are taken from \cite{sneden.2014ApJS..214...26S} and \cite{ram.2014ApJS..211....5R}.  We could not determine oxygen in both the stars as the lines [OI] 6300.3 and 6363.7 {\rm \AA} are highly blended and the near infrared oxygen triplet lines are out of the spectral coverage.

\subsection{Na, Mg, Ca, Sc, Ti}

The abundance of Na is estimated using the equivalent width measurements as well as the spectrum synthesis calculation of the lines Na I 5889.95 and 5895.92 {\rm \AA}. While sodium is found to be enhanced in HE 2155$-$2043 with [Na/Fe] $\sim$ 0.98, it is under abundant in HE~2148$-$2039 with [Na/Fe] $\sim$ $-0.32$. Abundance of Mg is determined using the measured equivalent widths as well as using the spectrum synthesis calculation of lines Mg I 5172.68 and 5183.60 {\rm \AA}. HE~2155$-$2043 exhibits enhancement of Mg with [Mg/Fe] $\sim$ 1.57. Magnesium is found to be under abundant in HE 2148$-$2039 with [Mg/Fe] $\sim$ $-0.35$. We have used the equivalent width measurements of three Ca I lines (Table \ref{table5}) to estimate the abundance of Ca in HE~2155$-$2043. For HE~2148$-$2039 the spectrum synthesis calculation of line Ca I 6162.17 is performed to determine the abundance of Ca. Calcium is slightly enhanced in HE~2155$-$2043, and near solar in HE~2148$-$2039. 

\par The abundance of Sc is determined using the equivalent width measurements of three Sc II lines (Table \ref{table5}) in HE~2155$-$2043. The spectrum synthesis calculation of Sc II 4320.73 {\rm \AA} is performed to estimate the abundance of Sc in HE~2148$-$2039. Scandium abundance is found to be near solar in both the programme stars with [Sc/Fe] $\sim$ $-0.19$ and $-0.02$ in HE~2148$-$2039 and HE~2155$-$2043 respectively. The abundance of Ti is measured using the equivalent width measurements of four Ti I lines (Table \ref{table5}). The abundance is nearly same in both the stars with a value [Ti/Fe] ${\sim}$  0.20.

\subsection{Cr, Co, Mn, Ni}

The abundances of Cr, Co, Mn and Ni could be determined only in 
HE~2155$-$2043. In HE~2148$-$2039, these lines are found to be blended and not useful for abundance determination. Abundance of Cr is determined using the spectrum synthesis calculation of Cr I 4289.72 {\rm \AA} that gives [Cr/Fe] $\sim$ $-0.10$. The spectrum synthesis calculation of Mn I 4030.75 {\rm \AA} gives a near solar value. The abundance of Co is determined using the spectrum synthesis calculation of Co I 4121.31 {\rm \AA} that gives [Co/Fe] $\sim$ 0.90. The spectrum synthesis of the line Ni I 5476.90 {\rm \AA} is used to derive the abundance of Ni. Nickel is only slightly enhanced in this object with [Ni/Fe] $\sim$ 0.30.

\subsection{Sr, Ba}

Among the neutron-capture elements, we could determine the abundances of only Sr and Ba in our programme stars. Spectrum synthesis of Sr II 4215.52 {\rm \AA} is used to estimate the abundance of Sr in HE~2148$-$2039 and is found to be under abundant with [Sr/Fe] $\sim$ $-2.02$. In this star, the Sr II 4077.71 {\rm \AA} line could not be detected. The spectrum synthesis calculation of these two lines give [Sr/Fe] $\sim$ $-0.04$ for HE~2155$-$2043.
\par
\begin{figure}
\centering
\includegraphics[width=12cm,height=11cm]{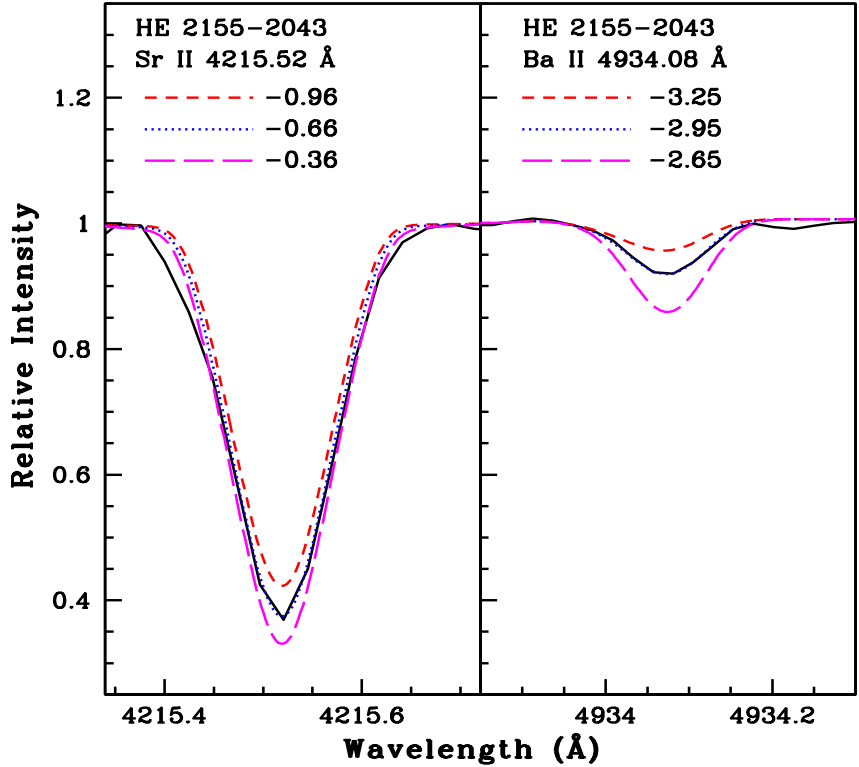}
\caption{Synthesis of Sr II line around 4215.52 {\rm \AA} and Ba II line around 4934 {\rm \AA}. Dotted line represents synthesized spectra and the solid line indicates the observed spectra. Short dashed line represents the synthetic spectra corresponding to $\Delta$ [X/Fe] = -0.3 and long dashed line  corresponds to $\Delta$[X/Fe] = +0.3}
\label{fig6}
\end{figure}

The abundance of Ba is estimated using  the spectrum synthesis calculation of  Ba II 4934.08 \AA\, in both the stars  considering the hyperfine structure and Ba isotopic ratios taken from \cite{McWilliam.1998AJ....115.1640M} and \cite{arlandini.1999ApJ...525..886A} respectively. The other prominent lines  Ba II 5853.66 and 6141.71 {\rm \AA}, are found to be extremely weak. Barium is under abundant in both the programme stars with [Ba/Fe] $\sim$ $-0.84$ and $-1.64$ for HE~2148$-$2039 and HE~2155$-$2043 respectively. The spectrum synthesis fits for Sr and Ba lines in HE~2155$-$2043 are shown in the Figure \ref{fig6}.
\par
We have checked for several absorption features due to other neutron-capture elements in the spectra of our program stars that are generally observed in extremely metal-poor stars. But these lines were found to be either absent or too weak to be used for  abundance determination. The abundance results for our programme stars are presented in Table \ref{table6}.

{\footnotesize
\begin{table*}
\caption{Elemental abundances in HE 2148$-$2039 and HE 2155$-$2043}
\resizebox{\textwidth}{!}{\begin{tabular}{|ccc|ccc|ccc|}
\hline
  &      &       & \multicolumn{3}{c}{HE~2148$-$2039} & \multicolumn{3}{c}{HE~2155$-$2043} \\
\hline
& Z & Solar log$\epsilon^{\ast}$ & log$\epsilon$& [X/H]& [X/Fe]&log$\epsilon$& [X/H]    &[X/Fe] \\ \hline
C (CH band)     & 6  &  8.43  & $<$ 6.28$\pm$0.20(syn)         & $<$ $-2.15$ & $<$ 1.15    & 7.05$\pm$0.20(syn)       & $-1.38$      & 2.05\\ 
N               & 7  &  7.83  & $<$ 8.39$\pm$0.20(syn)         & $<$ 0.56   & $<$ 3.86    & 6.50$\pm$0.20(syn)       & $-1.33$      & 2.10 \\
Na I	& 11 &	6.24 & 2.62$\pm$0.20(syn)         & $-3.62$ & $-0.32$ & 3.79$\pm$0.20(syn)         & $-2.45$      & 0.98 \\
Mg I	& 12 &	7.60 & 3.95$\pm$0.20(syn)         & $-$3.65 & $-0.35$ & 5.74$\pm$0.20(syn)         & $-1.86$      & 1.57 \\
Ca I	& 20 &	6.34 & 3.07$\pm$0.20(syn)         & $-$3.27 & 0.03    & 3.35$\pm$0.14(4)  & $-2.99$      & 0.44 \\
Sc II	& 21 &	3.15 & $-0.39$$\pm$0.20(syn)      & $-$3.54 & $-0.19$ & $-0.36$$\pm$0.07(3) & $-3.51$    & $-0.02$ \\
Ti II	& 22 &	4.95 & 1.82$\pm$0.06(3)  & $-$3.13 & 0.22    & 1.67$\pm$0.15(4)  & $-3.28$      & 0.21  \\
Cr I  & 24 &  5.64 & -                 &    -    &   -     & 2.11$\pm$0.20(syn)         & $-3.53$      & $-0.10$   \\ 
Mn I	& 25 &	5.43 &-                  & -       & -       & 2.04$\pm$0.20(syn)         & $-3.39$      & 0.04 \\
Fe I	& 26 &	7.50 &4.20$\pm$0.17(9)   & $-$3.30 & -       & 4.07$\pm$0.06(25) & $-3.43$      & -     \\
Fe II	& 26 &	7.50 &4.15(1)            & $-$3.35 & -       & 4.01$\pm$0.04(3)  & $-3.49$      & -    \\
Co I	& 27 &	4.99 & -                 &  -      & -       & 2.46$\pm$0.20(syn)         & $-2.53$      & 0.90   \\
Ni I	& 28 &	6.22 &-                  & -       & -       & 3.09$\pm$0.20(syn)         & $-3.13$      & 0.30  \\
Sr II & 38 &  2.87 &$-2.50$$\pm$0.20(syn)       & $-5.37$ & $-2.02$ & $-0.66$$\pm$0.20(syn)      & $-3.53$      & $-0.04$ \\
Ba II & 56 & 2.18  &$-2.01$$\pm$0.20(syn)       & $-4.19$ & $-0.84$ & $-2.95$$\pm$0.20(syn)      & $-5.13$      & $-1.64$  \\
\hline
\end{tabular}
\label{table6}
}

$\ast$  Asplund (2009), The number inside the  parenthesis shows the 
number of lines used for the abundance determination. 
\end{table*}
}

{\footnotesize
\begin{table*}
\caption{\bf Elemental abundances in HE~2155$-$2043 at log g = 0.40 and log g = 1.54}
\resizebox{\textwidth}{!}{\begin{tabular}{|ccc|ccc|ccc|c|}
\hline
  &      &       & \multicolumn{3}{c}{T$_{eff}$ = 4800, log g = 0.40, $\xi$ = 1.21} & \multicolumn{3}{c}{T$_{eff}$ = 4800, log g = 1.54, $\xi$ = 1.21}& \\     
\hline
& Z   & Solar log$\epsilon^{\ast}$ &log$\epsilon$& [X/H]    &[X/Fe] &log$\epsilon$& [X/H]    &[X/Fe]& Difference\\ \hline
C (CH band)     & 6  &  8.43  & 7.05$\pm$0.20(syn)       & $-1.38$      & 2.05 & 6.75$\pm$0.20(syn) & $-1.68$ & 1.66& 0.39\\ 
N               & 7  &  7.83  & 6.50$\pm$0.20(syn)       & $-1.33$      & 2.10 & 6.80$\pm$0.20(syn) & $-$1.03 & 2.31 & 0.21\\
Na I	& 11 &	6.24 & 3.79$\pm$0.20(syn)         & $-2.45$      & 0.98   & 3.83$\pm$0.20(syn) & $-$2.41 & 0.93 & 0.05\\
Mg I	& 12 &	7.60 & 5.74$\pm$0.20(syn)         & $-1.86$      & 1.57 & 5.44$\pm$0.20(syn) & $-$2.16 & 1.18& 0.39\\
Ca I	& 20 &	6.34 & 3.35$\pm$0.14(4)  & $-2.99$      & 0.44 & 3.27$\pm$0.14(4) & $-$3.07 & 0.27 & 0.17\\
Sc II	& 21 &	3.15 & $-0.36$$\pm$0.07(3) & $-3.51$    & $-0.02$ &  -0.11$\pm$0.07(3) & $-$3.26 & 0.09& 0.11\\
Ti II	& 22 &	4.95 & 1.67$\pm$0.15(4)  & $-3.28$      & 0.21& 1.88$\pm$0.15(4) & $-$3.07 & 0.28& 0.07 \\
Cr I  & 24 &  5.64 & 2.11$\pm$0.20(syn)         & $-3.53$      & $-0.10$ & 2.27$\pm$0.20(syn) & $-$3.37 & $-$0.03 & 0.07\\ 
Mn I	& 25 &	5.43 & 2.04$\pm$0.20(syn)         & $-3.39$      & 0.04 & 1.43$\pm$0.20(syn) & $-$4.00 & $-$0.66& 0.70\\
Fe I	& 26 &	7.50 & 4.07$\pm$0.06(25) & $-3.43$      & - & 4.16$\pm$0.06(25) & $-3.34$ & - & 0.09  \\
Fe II	& 26 &	7.50 & 4.01$\pm$0.04(3)  & $-3.49$      & - & 4.15$\pm$0.04(3) & $-3.35$ & -  & 0.14\\
Co I	& 27 &	4.99 & 2.46$\pm$0.20(syn)         & $-2.53$      & 0.90 & 2.18$\pm$0.20(syn) & $-$2.81 & 0.53 & 0.37\\
Ni I	& 28 &	6.22 & 3.09$\pm$0.20(syn)         & $-3.13$      & 0.30 & 3.00$\pm$0.20(syn) & $-$3.22 & 0.12 & 0.18\\
Sr II & 38 &  2.87 & $-0.66$$\pm$0.20(syn)      & $-3.53$      & $-0.04$ & $-$0.61$\pm$0.20(syn) & $-$3.48 & $-$0.13 & 0.09\\
Ba II & 56 & 2.18  & $-2.95$$\pm$0.20(syn)      & $-5.13$      & $-1.64$ & $-$2.68$\pm$0.20(syn) & $-$4.86 & $-$1.51 & 0.13\\
\hline
\end{tabular}
}

$\ast$  \cite{asplund.2009ARA&A..47..481A}, The number inside the  parenthesis shows the 
number of lines used for the abundance determination.\\ 
Column 10 represents the difference in the abundances obtained for log g = 0.40 and log g = 1.54
\label{table7}
\end{table*}
}

\section{Discussion}\label{sec:discussion}
\begin{figure}
\centering
\includegraphics[width=12cm,height=11cm]{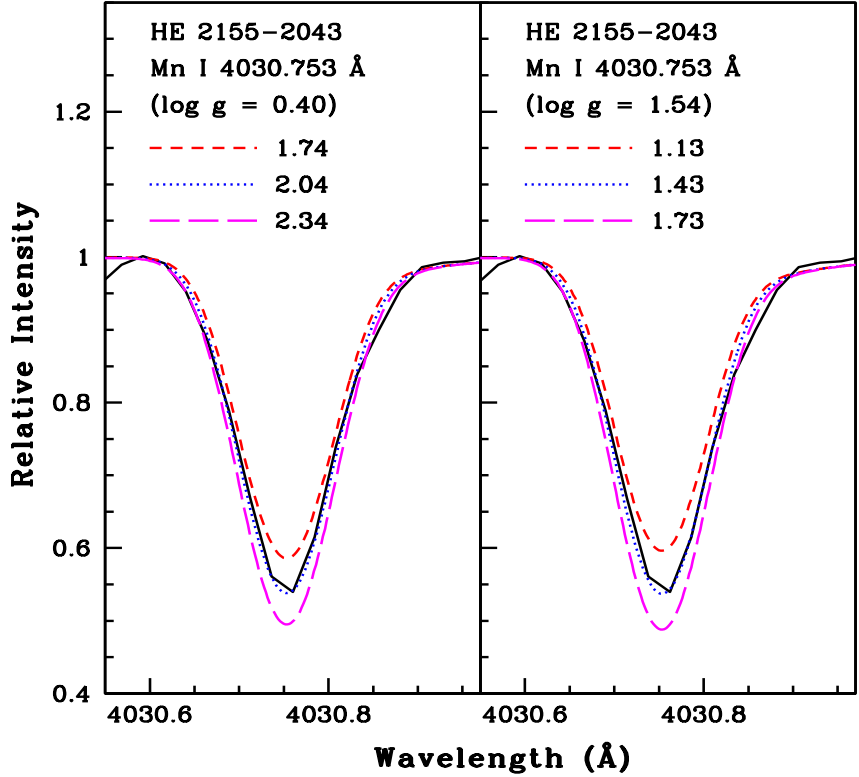}
\caption{Synthesis of Mn I line around 4030.75 {\rm \AA} for log g = 0.40 and 1.54. Dotted line represents synthesized spectra and the solid line indicates the observed spectra. Short dashed line represents the synthetic spectra corresponding to $\Delta$[Mn/Fe] = -0.3 and long dashed line  corresponds  to $\Delta$[Mn/Fe] = +0.3}
\label{fig7}
\end{figure}

Before we present a detailed discussion and interpretation of the abundance results obtained from the spectroscopic analysis for the two programme stars, we would like to comment on the adopted surface gravity value for the star HE~2155$-$2043. The value of log g for which Fe II lines give the same abundance as obtained from the Fe I lines, is taken as the spectroscopic log g  value of the star and in general, the spectroscopic log g match with their counter parts obtained from the parallax method within the errors. 
For the object HE~2155$-$2043, the three Fe II lines  used return a value 0.4 for log g. This value is very different from the log g (1.54) as obtained from the parallax method. We have therefore calculated the abundances for all the elements at both the values of log g keeping the spectroscopic temperature and the corresponding microturbulent velocity constant. The abundance obtained using log g = 1.54 are presented in Table \ref{table7}. The observed difference may be due to the high carbon enhancement in the object. For a star which shows strong enhancement in carbon, the evolutionary track is shifted towards lower temperatures which causes inconsistency between the atmospheric parameters determined from the evolutionary track and that derived from the spectral analysis as the exact location of the giant branch is sensitive to the opacity in the atmosphere of a star \citep{marigo.2002A&A...387..507M,jorrisen.2016A&A...586A.159J}. The error in parallax for HE 2155$-$2043 is around 38\% which can also contribute towards the observed deviation. The difference in the abundance values for various elements obtained using the log g from parallax and that from spectroscopy ranges from 0.05 to 0.70 dex. However the classification of HE~2155$-$2043 as CEMP-no star remains the same. An example for the spectrum synthesis of Mn for log g = 1.54 and log g = 0.40 is shown in Figure \ref{fig7}.
\subsection{Locations of the programme stars in the A(C) vs. [Fe/H] diagram and classification:}
The existance of different carbon production mechanisms among CEMP stars were explored by many authors.  \cite{spite.2013} identified the existance of a bimodality among the CEMP stars in a plot of absolute carbon abundance as a function of metallicity, based on a sample of $\sim$ 50 stars. They showed that the stars in their sample with metallicity $>$ $-3.0$ occupy the region around A(C) $\sim$ 8.25 which they have taken as higher carbon "plateau". Objects that fall around this band are mostly composed of CEMP-s and CEMP-r/s stars. The stars with [Fe/H] $<$ $-3.4$ occupy a region around A(C) $\sim$ 6.5 and found to be dominated by CEMP-no stars. These observations led to their interpretation that two different carbon production mechanisms exist among CEMP stars. \cite{bonifacio.2015A&A...579A..28B} and \cite{hansen.2015ApJ...807..173H} confirmed this bimodality based on even larger sample of stars. The knowledge of binary status of the stars of the two different carbon bands may provide important clues regarding the nature of their progenitors.

\begin{figure}
\centering
\includegraphics[width=12cm,height=11cm]{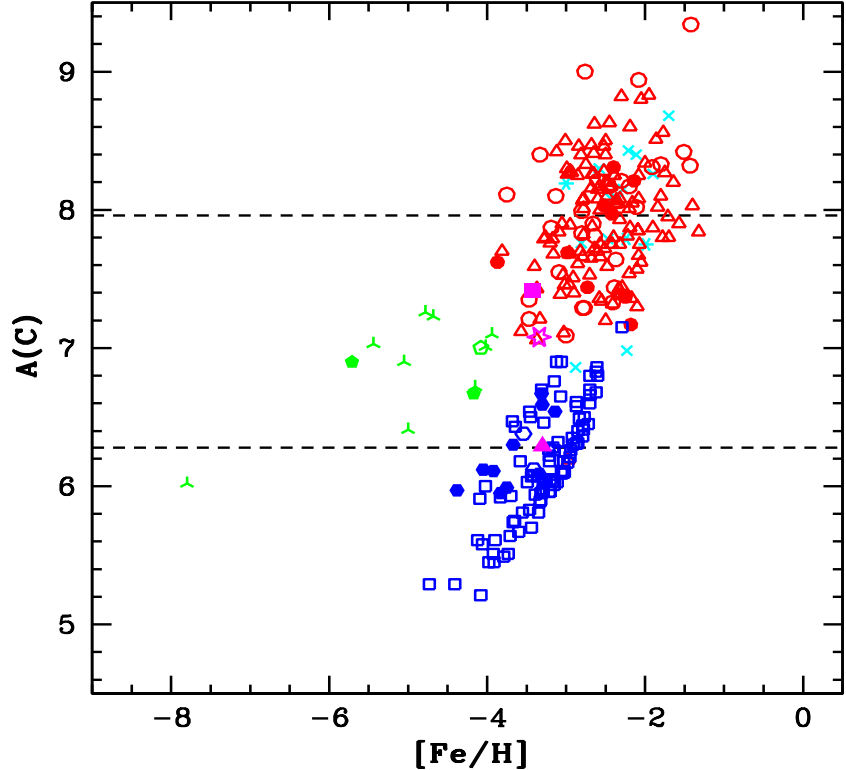}
\caption{A(C) vs. [Fe/H] diagram for CEMP stars compiled  by  \cite{yoon.2016ApJ...833...20Y}. In this plot Cyan colour indicates CEMP-r/s stars: binary stars are represented by eight sided star and stars with no information on  their binary status are indicated by cross symbols. Open and filled red circles indicate binary and single CEMP-s stars respectively. CEMP-s stars with no information on their binarity are represented using red open triangles.  Open and filled hexagons in blue  represent binary and single CEMP-no stars respectively. CEMP-no stars with no information on their binarity are indicated  by blue open squares. Group III CEMP-no stars are represented using green symbols: binary and single stars are represented by open and filled pentagons respectively. Skeleton triangles represent stars with no information about binary status. Programme stars are represented using magenta colour, HE~2148$-$2039 is represented by a filled triangle and HE~2155$-$2043 is represented by a filled square.}
\label{fig8}
\end{figure}

Location of a carbon  star in   the absolute carbon abundance A(C) vs. metallicity [Fe/H]  diagram \citep{yoon.2016ApJ...833...20Y} could be an important indicator of its class. We have therefore attempted to classify the programme stars from their locations in such a plot (Figure \ref{fig8}). \cite{yoon.2016ApJ...833...20Y} compiled high-resolution  abundance data for 305 CEMP stars  and found that the distribution of  these CEMP stars in the A(C) vs. [Fe/H] diagram  is bimodal  with two distinct peaks, the high-carbon region   centered around A(C)  = 7.96 and the low-carbon region  centered  around A(C) = 6.28. The CEMP-s and CEMP-r/s stars are found to  occupy the high-carbon region and are called as the Group I objects. The majority of CEMP-no stars are found to occupy the low-carbon region. Among the CEMP-no  stars, the group that show a clear trend with respect to [Fe/H] is called the group II objects, and those that do not show any trend with respect to [Fe/H] are classified as Group III objects. The Group III and the lower end of Group I has a very thin separation. It is in this region the object HE~2155$-$2043 lies close to the lower boundary of the region for Group I, and the upper boundary for the  Group III objects  making it difficult to clearly classify this object based on this diagram alone.  However, on the basis of its very low barium abundance [Ba/Fe] ${\sim}$ $-$1.6,  the object can be  classified as a CEMP-no star. It is to be noted that, in general the CEMP-s and CEMP-r/s stars show [Ba/Fe] ${\ge}$ 1.0.      
We have applied the necessary  correction factors \citep{placco.2014bApJ...797...21P} to our estimated carbon abundances of the programme stars using the public online tool \footnote{\url{http://vplacco.pythonanywhere.com/}} for placing them in 
Figure \ref{fig8}. The corrections applied to the estimated carbon abundances are  0.01 and  0.37 for HE~2148$-$2039 and HE 2155$-$2043 respectively.  It is clearly seen from the figure  that the object HE~2148$-$2039 could be a bonafied CEMP-no star  as it  falls in the region occupied by CEMP-no Group II objects.

\subsection{Analysis of the observed abundance patterns in the program stars}

Understanding how the CEMP-no stars become enriched in various elements can provide insight into their nucleosynthetic pathways and the formation history. The estimated abundances of alpha, Fe-peak and the neutron-capture elements such as Sr and Ba in the two programme stars match well with the observed abundance patterns of these elements in other extremely metal-poor objects reported by various studies. Carbon and nitrogen are found to be enhanced in both the programme stars. HE 2148$-$2039 shows very high nitrogen abundance with [N/Fe] $\sim$ 3.87. Such a high value of nitrogen abundance is not uncommon among extremely metal-poor objects. \cite{roederer.2014AJ....147..136R} reported an enhancement in nitrogen with [N/Fe] $>$ 3.0 for the objects 
CS~22171$-$037, CS~22177$-$009, CS~22884$-$108 and CS~22960$-$053 with similar metalicities.

\par
\begin{figure}
\centering
\includegraphics[width=12cm,height=10cm]{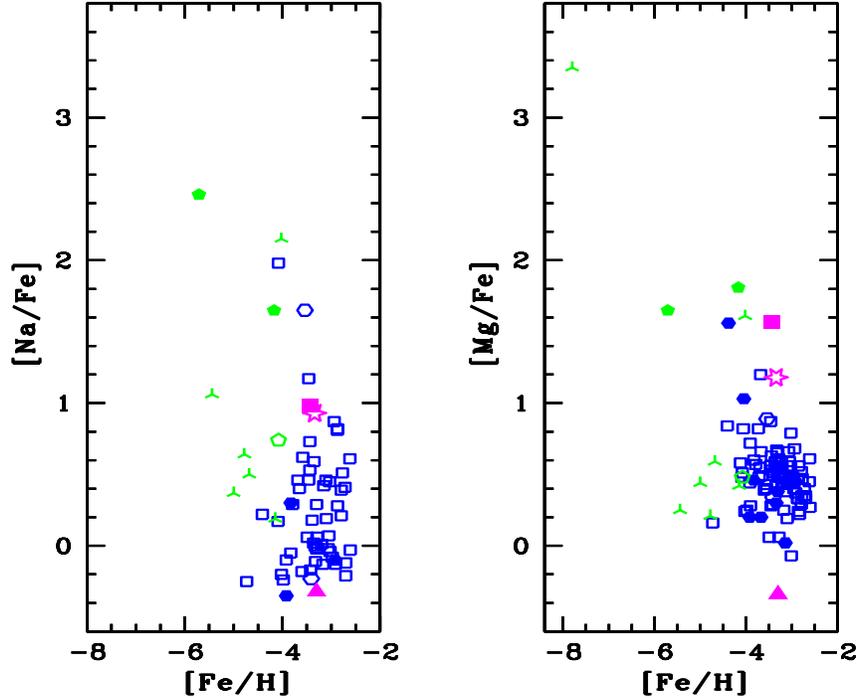}
\caption{Distribution of the [Na/Fe] and [Mg/Fe] as a function of metallicity for CEMP-no stars. The data used in the plot are taken from Yoon et al. (2016). Symbols used are same as in Figure \ref{fig8}}
\label{fig9}
\end{figure}

\begin{figure}
\centering
\includegraphics[width=12cm,height=10cm]{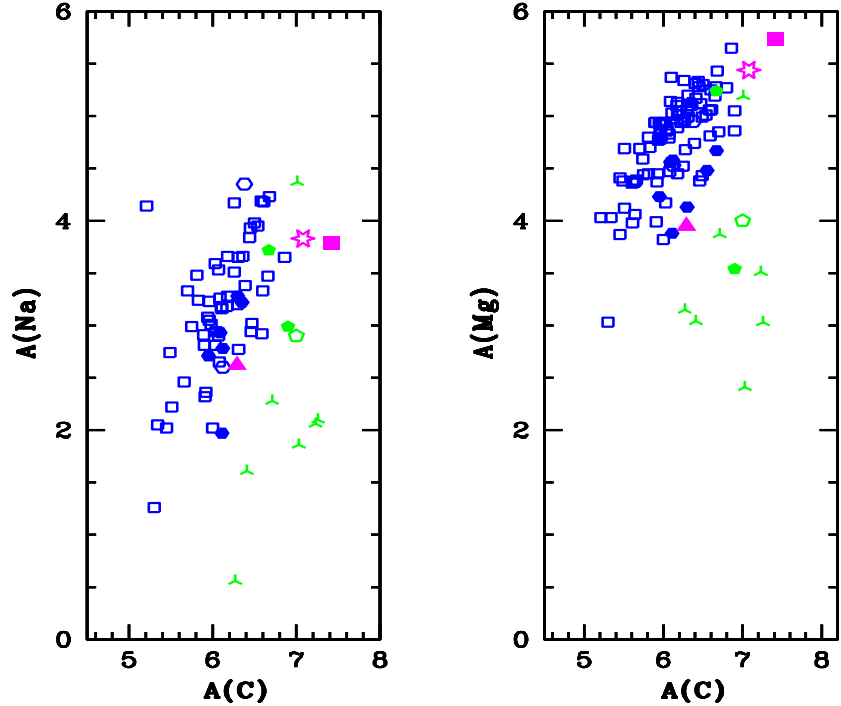}
\caption{Distribution of the A(Na) and A(Mg) as a function of A(C) for CEMP-no stars. The data used in the plot are taken from Yoon et al. (2016). Symbols used are same as in Figure \ref{fig8}.}
\label{fig10}
\end{figure}

  
Our programme stars exhibit light element abundance patterns that are distinctly different from each other indicating different production  mechanisms for these two stars. While HE~2155$-$2043 shows enhancement of Na and Mg, these elements are underabundant in HE 2148$-$2039. Previous studies show that the enhancement of light elements such as Na, Mg, Al, and Si are common among the majority of the extremely metal-poor stars \citep{aoki.2002ApJ...576L.141A,frebel.2005Natur.434..871F}. For example, \cite{roederer.2014AJ....147..136R} reported such a high enhancement in Na and Mg for the extremely metal-poor stars CS 29498-043, CS 22893-010 and HE 1012-1540. \cite{aoki.2002ApJ...576L.141A} and \cite{masseron.2010A&A...509A..93M} divided the CEMP-no stars into two categories with Mg- enhanced and Mg- normal at extremely low metallicities. \cite{yoon.2016ApJ...833...20Y} show that CEMP-no Group III stars exhibit enhancement of Na as well as Mg unlike CEMP-no Group II stars that do not show any enhancement of these elements. The locations of our programme stars in the [Na/Fe] vs. [Fe/H] and [Mg/Fe] vs. [Fe/H] (Figure \ref{fig9}) show that the programme star HE 2148$-$2039 is a  CEMP-no Group II object and HE 2155$-$2043 is a CEMP-no Group III object. Their locations in the A(Na) vs. A(C) and A(Mg) vs. A(C) (Figure \ref{fig10}) also support the same. This again shows the existance of different progenitors among the CEMP-no stars themselves, 
leading to two different groups.

\par
\begin{figure}
\centering
\includegraphics[width=12cm,height=10cm]{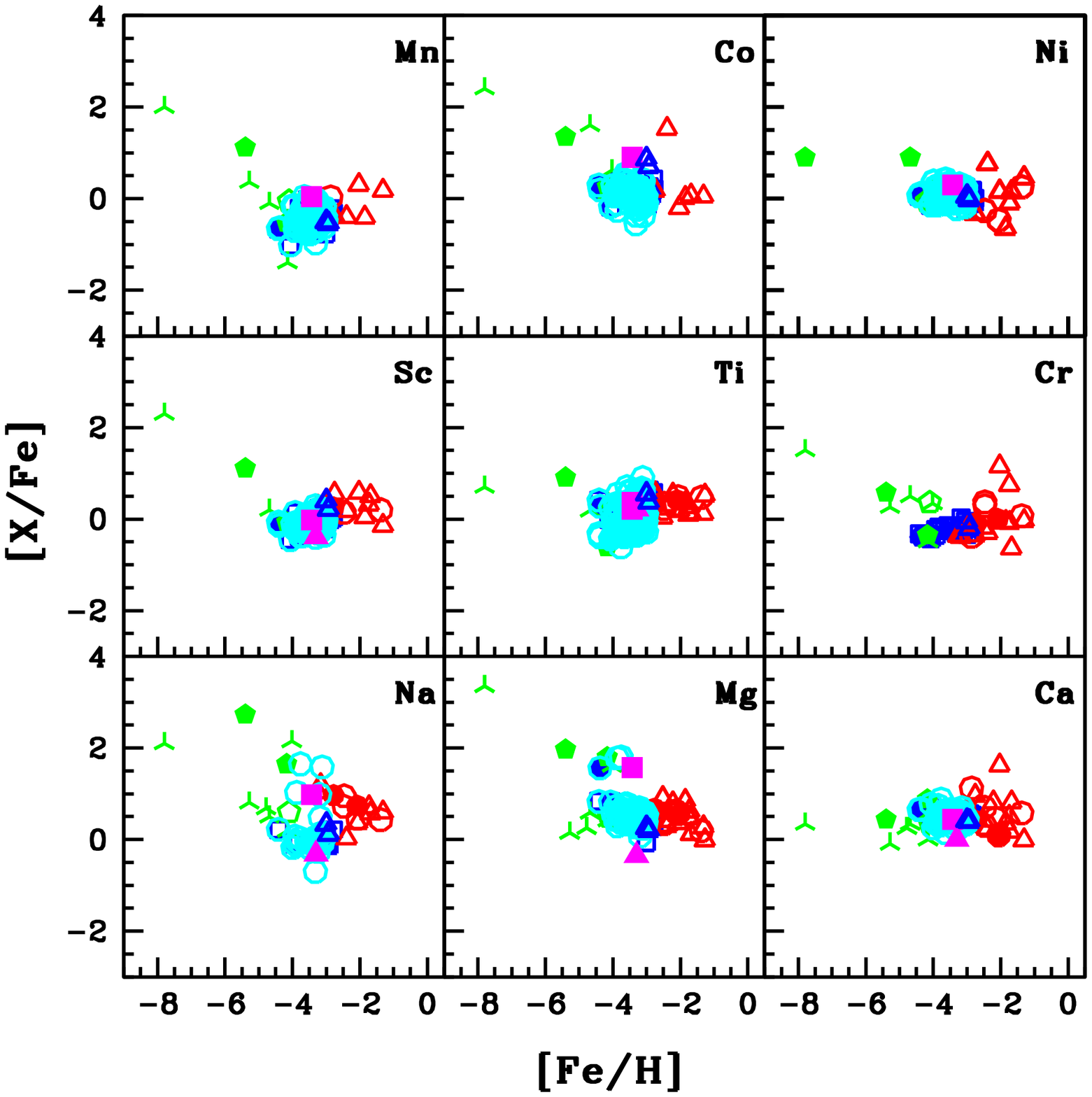}
\caption{Abundance ratios of light elements with respect to metallicity estimated in our programme stars. Symbols used are same as in Figure 8. The abundance values for the CEMP-no stars are taken from \cite{christlieb.2004A&A...428.1027C},\cite{plez.cohen.2005A&A...434.1117P},\cite{yong.2013ApJ...762...26Y}, \cite{hansen.2014ApJ...787..162H}, \cite{bonifacio.2015A&A...579A..28B}, \cite{bessel.2015ApJ...806L..16B} and \cite{frebel2018}. The abundance values for CEMP-s stars are taken from \cite{lucatello.2003AJ....125..875L}, \cite{barklem.2005A&A...439..129B}, \cite{cohen.2006AJ....132..137C}, \cite{goswami.2006MNRAS.372..343G}, \cite{aoki.2007ApJ...655..492A}, \cite{karinkuzhi.2015MNRAS.446.2348K}, \cite{purandardas.2019bBSRSL..88..207P} and \cite{purandardas.2019aMNRAS.486.3266P}. Abundances of light elements observed in the programme stars are also compared with the abundance patterns of extremely metal-poor objects (cyan open circles) presented by \cite{roederer.2014AJ....147..136R} and with the extremely metal-poor stars in the Sylgr stellar stream (blue open triangle) presented by \cite{roaderer.2019ApJ...883...84R} .  }
\label{fig11}
\end{figure}

\begin{figure}
\centering
\includegraphics[width=10cm,height=10cm]{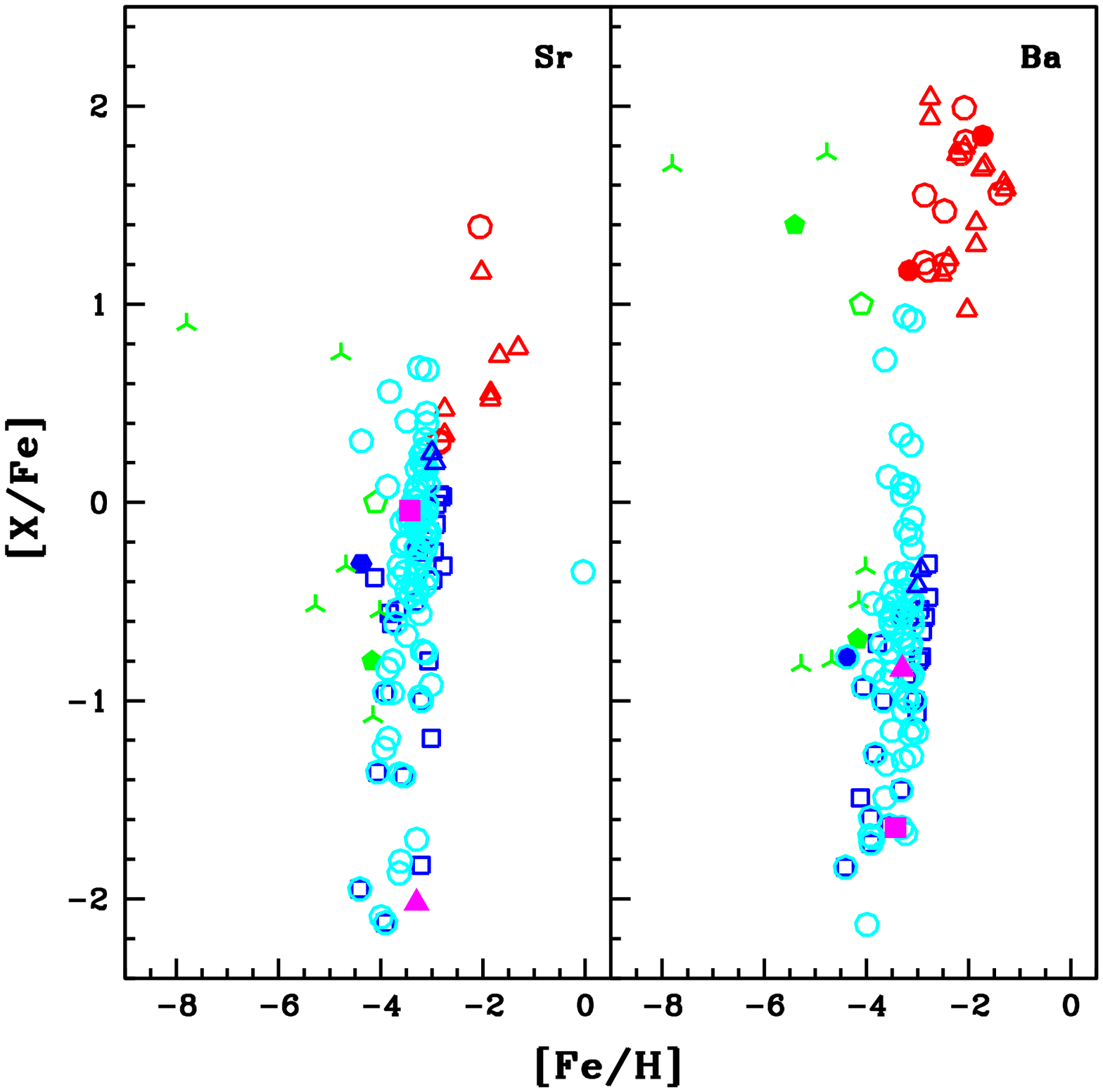}
\caption{ Same as Figure 8 but for Sr and Ba.}
\label{fig12}
\end{figure}

 In Figures \ref{fig11} and \ref{fig12},  we have  shown  a comparison between the estimated elemental abundance ratios of  the  programme stars  with their counterparts observed in CEMP-s, CEMP-no and  extremely metal-poor stars taken from various sources in literature.   The estimated abundance ratios are also  compared  with their counterparts observed in  two extremely metal poor CEMP-no stars  J1202-0020 and J1208-0029 of the Sylgr stellar streams   \citep{roaderer.2019ApJ...883...84R} in the same figures.  It is clear from the figures that the abundance ratios of the program stars match well with those observed in the  literature CEMP-no stars.
\par While the  carbon is more enhanced in the two programme stars than the Sylgr stellar stream CEMP-no stars,  nitrogen is found to have  similar values. The ${\alpha}$-elements in  HE 2155$-$2043 are found to be  over abundant and in HE 2148$-$2039 under abundant when compared with those observed in the two Sylgr stellar stream stars.  While the abundances of Fe-peak elements in HE~2155$-$2043 are found to be similar to that of Sylgr stellar stream stars, the estimated abundances of Sr and Ba in the two programme stars are found to be lower.

\subsection{ Possible progenitors of the program stars}
 \cite{Frischknecht.2012A&A...538L...2F} studied the impact of rotation-induced mixing on the neutron capture nucleosynthesis in rotating massive stars at low metallicities using the Geneva stellar evolution code. Their studies show that the s-process nucleosynthesis in massive stars produces elements upto Sr at a metallicity, [Fe/H] $>$ $-2$. They obtained [Sr/Ba] $>$ 0 from low metallicity massive rotating stars. \cite{choplin.2017A&A...607L...3C} also showed that the [Sr/Ba] ratio is higher in fast rotating massive stars than in AGB stars. Since [Sr/Ba] ratio is different in AGB and fast rotating massive stars, it is a useful descriptor to trace the formation sites \citep{hansen.2019A&A...623A.128H}. The estimated [Sr/Ba] ratio in HE 2155$-$2043 is $\sim$ 1.60. This implies that the possible progenitor of this object may be a fast rotating massive star. 
\par  \cite{yoon.2016ApJ...833...20Y} suspect that the CEMP-no Group III objects may be associated with spin star progenitors and Group II objects may be related to the faint supernovae progenitors. However they remain open to other classes of progenitors that may also contribute to the observed abundance anomalies. The [Sr/Ba] ratio in HE~2148$-$2039 is $\sim$ $-1.18$. [Sr/Ba] ratio is found to be less than 0 for metal-poor AGB stars \citep{cristallo.2011ApJS..197...17C,hansen.2019A&A...623A.128H}. This may point towards a metal-poor AGB progenitor for HE~2148$-$2039. According to \cite{arentsen.2019A&A...621A.108A}, CEMP-no stars can be in a binary system. The primary companion is an extremely metal-poor star that once passed through the AGB phase and it had not produced any significant amount of s-process elements.

\par
\begin{figure}
\centering
\includegraphics[width=11cm,height=11cm]{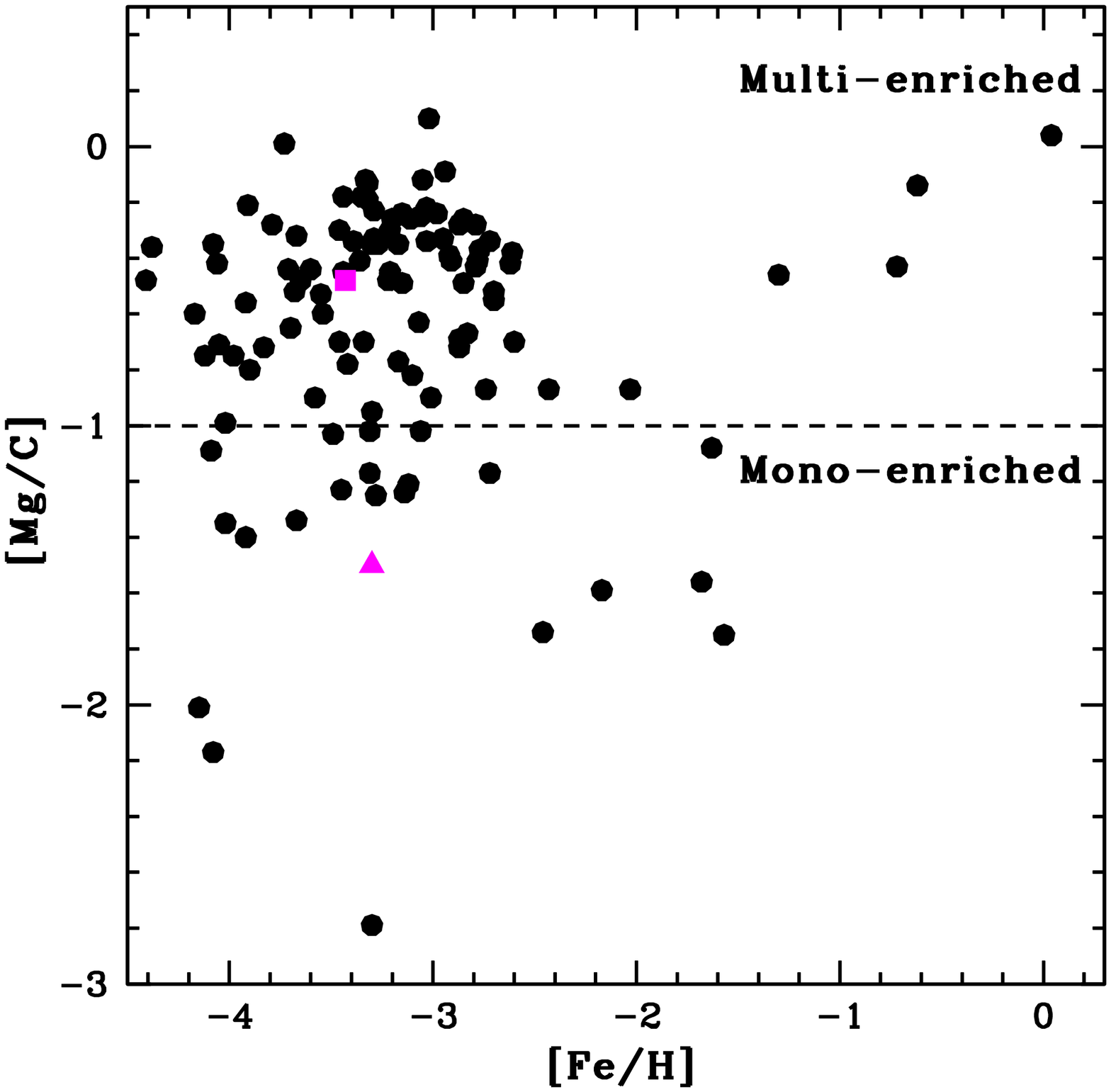}
\caption{Location of programme stars in [Mg/C] vs. [Fe/H] plot. Filled triangle represents HE~2148$-$2039 and filled square represents HE~2155$-$2043. Filled circles represent the CEMP stars from literature \citep{norris.2007ApJ...670..774N,hollek.2011ApJ...742...54H,ito.2013ApJ...773...33I,yong.2013ApJ...762...26Y,cohen.2013ApJ...778...56C,hansen.2014ApJ...787..162H,placco.2014bApJ...797...21P,roederer.2014AJ....147..136R,placco.2014bApJ...797...21P,hansen.2015ApJ...807..173H,li.2015PASJ...67...84L,jacobson.2015ApJ...807..171J,bessel.2015ApJ...806L..16B,karinkuzhi.2015MNRAS.446.2348K, purandardas.2019bBSRSL..88..207P,purandardas.2019aMNRAS.486.3266P,partha.2021arXiv210109518G,shejeela.2021MNRAS.502.1008S}}
\label{fig13}
\end{figure}

 \cite{hartwig.2018MNRAS.478.1795H} showed that [Mg/C] ratio can be used to tell if a CEMP star was enriched by a single event (mono-enriched) or several pollution events (multi-enriched) based on their model calculations. For mono-enriched objects, [Mg/C] is found to be less than $-1.0$. Other elemental ratios such as [Sc/Mn] $<$ 0.5, [C/Cr] $>$ 0.5 and [Ca/Fe] $>$ 2.0 can also indicate mono-enrichment \citep{hartwig.2018MNRAS.478.1795H}. The estimated [Mg/C] ratio in our programme stars show that HE~2148$-$2039 is mono-enriched and HE~2155$-$2043 is multi-enriched (Figure \ref{fig13}). But the observed [Ca/Fe] in HE~2148$-$2039 and HE~2155$-$2043 do not support mono-enrichment. The [Sc/Mn] and [C/Cr] ratios estimated in HE~2155$-$2043 show that this object is enriched by a single pollution event.

\subsection{Mixing diagnostic}
\par
\begin{figure}
\centering
\includegraphics[width=11cm,height=11cm]{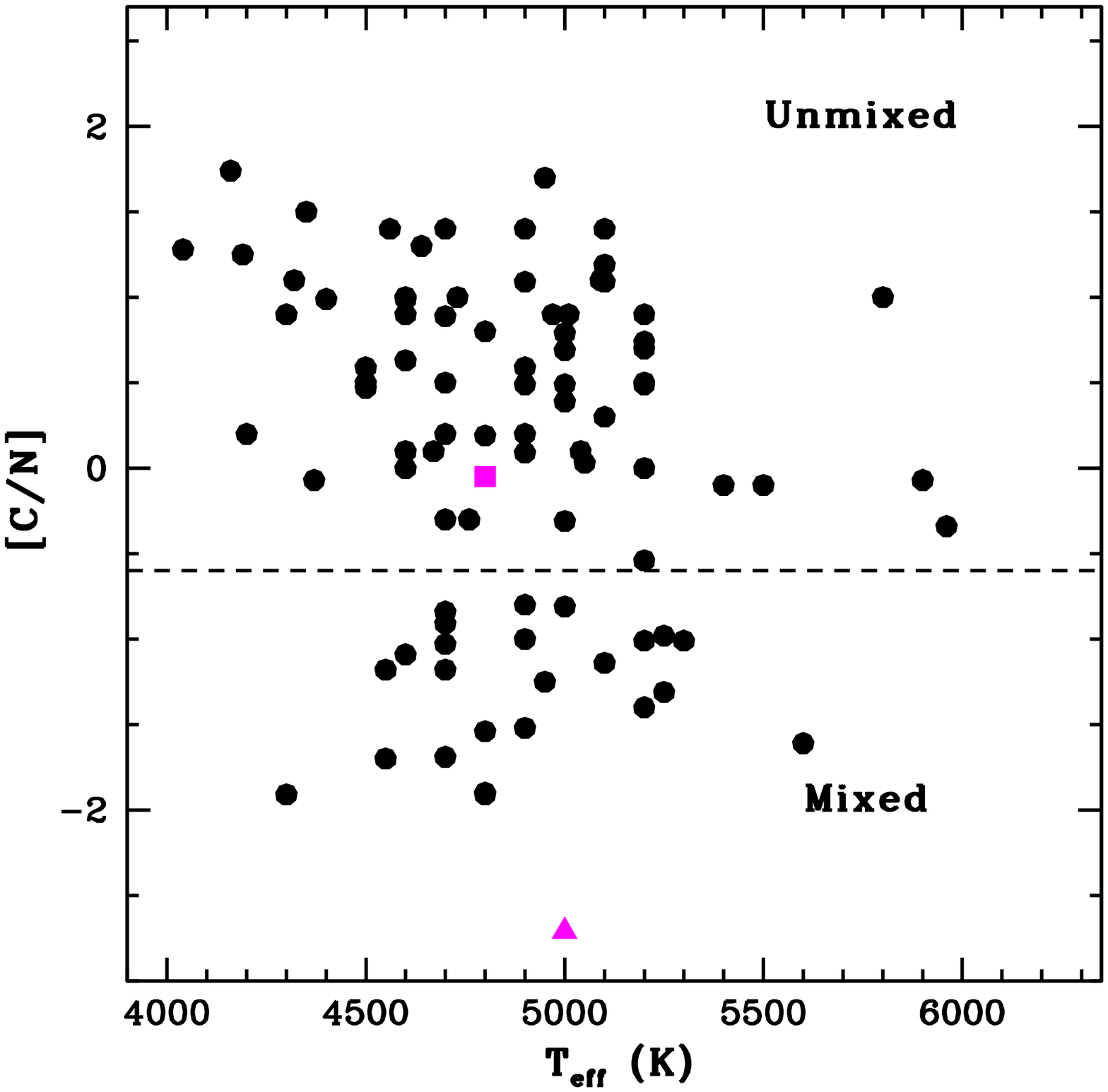}
\caption{Position of the programme stars in the [C/N] vs. T$_{eff}$ diagram. Filled triangle represents HE~2148$-$2039 and filled square represents HE~2155$-$2043. Filled circles represent the CEMP stars from literature \citep{spite.2006A&A...455..291S,aoki.2007ApJ...655..492A,hansen.2014ApJ...787..162H,goswami.2016MNRAS.455..402G,purandardas.2019bBSRSL..88..207P,purandardas.2019aMNRAS.486.3266P,partha.2021arXiv210109518G,shejeela.2021MNRAS.502.1008S}}
\label{fig14}
\end{figure}

 Various mixing processes have been found to operate in stars \citep{spite.2006A&A...455..291S, gratton.2000A&A...354..169G}. It is therefore important to understand whether the stars have undergone any internal mixing. [C/N] ratio can give important clues about the internal mixing process in a star. We have checked for the signatures of any internal mixing in our programme stars based on the estimated [C/N] ratios (Figure \ref{fig14}). Stars with [C/N] $>$ $-0.60$ are found to be unmixed stars \citep{spite.2005A&A...430..655S}. While HE 2148$-$2039 shows signatures of internal mixing, HE~2155$-$2043 falls in the unmixed region with [C/N] = $-0.05$. An extra mixing is found to operate in stars with luminosity, log(L/L$_{\odot}$) $\sim$ 2.0 \citep{gratton.2000A&A...354..169G, spite.2005A&A...430..655S}. Since HE~2155$-$2043 has log(L/L$_{\odot}$) $>$ 2.0, we expect signatures of mixing in this object. But this object is found to fall in the unmixed category based on [C/N] ratio. According to \cite{spite.2006A&A...455..291S}, [C/N] ratio is not a clean indicator of mixing, because, the abundances of carbon and nitrogen in the interstellar medium from which these stars are formed show large variations. But carbon isotopic ratio can be used as a good indicator of mixing, since it is high in primordial matter ($^{12}$C/$^{13}$C $>$ 70) and it is insensitive to the choice of atmospheric parameters for the stars (Spite et al. 2006). However we could not  estimate carbon isotopic ratio in our programme stars as the C$_{2}$ swan band around 4700 {\rm \AA} is found to be absent.

\section{Kinematic analysis}\label{sec:kinematic-analysis}

We have determined the space velocities of the programme stars following the methods as given in \cite{bensby.2003A&A...410..527B}. The space velocity of the star with respect to the local standard of rest (LSR) is given by\\
\begin{center}
  $(U, V, W)_{LSR} =(U,V,W)+(U, V, W)_{\odot}$ km/s.
  \end{center}

 where, $(U, V, W)_{\odot} =(11.1, 12.2, 7.3)$ km/s \cite{schonrich.2010MNRAS.403.1829S} is the solar motion with respect to LSR , and

\[\left[\begin{array}{lll}
 & U & \\
 & V & \\
 & W & 
\end{array} \right] = B\left[\begin{array}{lll}
  & k\mu_{\alpha }/\pi   & \\
 & k\mu_{\delta }/\pi  & \\
& \rho  & 
\end{array} \right]\]
where B = T.A, T represents the transformation matrix connecting Galactic coordinates and equatorial coordinates which is given by\\
 \[T = \left[\begin{array}{ccc}
-0.0548756 & -0.8734371 & -0.4838350 \\
+0.4941094  & -0.4448296 & +0.7469822  \\
-0.8676661 & -0.1980764 & +0.4559838 
\end{array} \right]\]
A represents the coordinate matrix which is defined as:\\

 \[A = \left[\begin{array}{ccc}
-sin\alpha  & -cos\alpha sin\delta  & +cos\alpha cos\delta  \\
+cos\alpha    & -sin\alpha sin\delta  & sin\alpha cos\delta   \\
0  & +cos\delta & sin\delta 
\end{array} \right]\]
 Where $\alpha$ is the right ascension and $\delta$ is the declination in degrees, k = 4.74057 km s$^{-1}$, $\mu _{\alpha }$ is the proper motion 
in right ascension, in arcsec yr$^{-1}$, $\mu _{\delta }$ is the proper motion in declination, in arcsec yr$^{-1}$, $\rho$ is the radial velocity in kms$^{-1}$  and $\pi$ is the parallax in arcsec. The proper motion and the parallax of the star are taken from (\cite{gaia.2016A&A...595A...1G}, \cite{gaia.2018A&A...616A...1G}) and the radial velocity is taken to be the value which we have estimated. The coordinate system that we used for U, V and W is right handed and it is therefore, they are positive in the directions of the Galactic center, Galactic rotation, and the North Galactic Pole (NGP) respectively \citep{johnson&soderblom.1987AJ.....93..864J}. The total spacial velocity of the star is 

\[ V_{spa}^{2}=U_{LSR}^{2}+V_{LSR}^{2}+W_{LSR}^{2} \]

\par We have also determined the membership of the star into the thin disc, thick disc or the halo population. We have followed the procedures as given in \cite{reddy.2006MNRAS.367.1329R}, \cite{bensby.2003A&A...410..527B,bensby.2004A&A...415..155B}, \cite{mishenina.2004A&A...418..551M} based on the assumption that the Galactic space velocity of the star has Gaussian distribution:

\[f(U,V,W) = K\times\exp[-\frac{U_{LSR}^{2}}{2\sigma _{U}^{2}}-\frac{(V_{LSR}-V_{asy})^{2}}{2\sigma _{V}^{2}}-\frac{W_{LSR}^{2}}{2\sigma _{W}^{2}}] \]

Where $K = \frac{1}{(2\pi) ^\frac{3}{2}\sigma _{U}\sigma _{V}\sigma _{W}}$\\
The values for characteristic velocity dispersion $\sigma _{U}$, $\sigma _{V}$  and $\sigma _{W}$ and the asymmetric drift $V_{asy}$ are adopted from \cite{reddy.2006MNRAS.367.1329R}. The results of the kinematic analysis of the programme stars are presented in Table \ref{table8}. Both the programme stars are found to be the members of the halo of the Galaxy. This is in agreement with the studies by \cite{carollo.2014ApJ...788..180C} that the CEMP-no stars are mostly found to be associated with the outer-halo population and the CEMP-s and CEMP-r/s stars are mostly found in the inner-halo. 
 \cite{tumilson.2007ApJ...665.1361T} suggests that the dominant progenitors of the CEMP population in the outer-halo are more massive stars than that for the inner-halo population for which the intermediate-mass stars might be the possible progenitors.

{\footnotesize
\begin{table*}
\caption{\bf Spatial velocity and probability estimates for 
the programme stars}
\begin{tabular}{lcccccccc}
\hline                       
Star name           & U$_{LSR}$         & V$_{LSR}$           & W$_{LSR}$ & V$_{spa}$  & p$_{thin}$ & p$_{thick}$ & p$_{halo}$ & Population \\
                    & (kms$^{-1}$)        & (kms$^{-1}$)          & (kms$^{-1}$) &  (kms$^{-1}$) &           &           & &  \\  
    \hline
HE~2148$-$2039   & $-112.80$$\pm$31.69 & $-109.10$$\pm$30.34 & $-163.95$$\pm$38.32 & 226.95$\pm$58 & 0.00 & 0.08 & 0.91 & Halo \\
HE~2155$-$2043   & $-192.29$$\pm$71.87 & $-442.13$$\pm$186.01 & $-253.59$$\pm$135.62 & 544.76$\pm$239.06 & 0.00 & 0.00 & 1.00 & Halo \\
\hline
\end{tabular}
\label{table8}
\end{table*}
}

\section{Conclusion}\label{sec:conclusion}
We have presented the results of the high-resolution spectroscopic analysis of two stars HE~2148$-$2039 and HE~2155$-$2043. Our analysis shows that the objects are extremely metal-poor with [Fe/H] $\sim$ $-3.30$ and $-3.43$ for HE~2148$-$2039 and HE~2155$-$2043 respectively. The observed abundances of carbon, A(C) = 6.29 and 7.42 (after applying necessary corrections) and the estimated [Ba/Fe] of $-0.84$ and $-1.64$ for HE~2148$-$2039 and HE~2155$-$2043 respectively confirm the programme stars to be CEMP-no stars. 
\par Among the neutron-capture elements, abundances of only Sr and Ba could be determined in our programme stars. Strontium is found to be under-abundant in HE~2148$-$2039 and is near solar in HE~2155$-$2043. The [Sr/Ba] ratio of HE~2155$-$2043 indicates that the possible progenitor for this object could be a fast rotating massive metal-poor star. [Sr/Ba] = $-1.18$ for HE~2148$-$2039 indicates a metal-poor AGB progenitor for this object. The lines due to other neutron-capture elements are found to be either absent or too weak to measure. 
\par The abundances of Na, and Mg estimated in our programme stars show that HE~2148$-$2039 is a CEMP-no Group II object and HE~2155$-$2043 is a CEMP-no Group III object as per the classification scheme of \cite{yoon.2016ApJ...833...20Y}. While the estimated abundance ratios of [Sc/Mn] and [C/Cr] in HE~2155$-$2043 show that this object is enriched by a single pollution event, the estimated [Mg/C] ratio indicates that the object is enriched by multiple pollution events. As we could not determine the abundances of Sc, Mn and Cr in 
HE~2148$-$2039, the formation history of this object could not be constrained based on these abundances. But the estimated [Mg/C] ratio of this object shows that this is a mono-enriched object. 
\par [C/N] ratio can act as an indicator of mixing. Locations of the two objects in the [C/N] vs. T$_{eff}$ plot show that the object HE 2148$-$2039 had undergone significant internal mixing and HE~2155$-$2043 does not show any signatures of mixing. Kinematic analysis shows that both the programme stars belong to the halo population of the Galaxy.
\par Low-mass extremely metal-poor stars are ideal candidates to study the nature of the very first  stars  as these objects are believed  to  bear the chemical imprints of the oldest metal-poor stars. However,   most of the extremely metal-poor stars exhibit diversity in  elemental abundance patterns  indicating different formation mechanisms.  Although there have been many studies,  the formation mechanisms of these stars and  their diverse abundance patterns are far from being completely understood.   The high-resolution  spectroscopic studies available for these stars are also currently limited in number. This underscores the need  for high resolution spectroscopic studies for  larger sample  that would provide robust observational constraints for theoretical studies. 

\section{ACKNOWLEDGEMENT}
\noindent Funding from the DST SERB project EMR/2016/005283 is gratefully acknowledged.   We thank the referee for many constructive suggestions and useful  comments on the manuscript  which improved this paper.   This work made use of the SIMBAD astronomical database, operated at CDS, Strasbourg, France, the NASA ADS, USA and data from the European Space Agency (ESA) mission Gaia (\url{https://www.cosmos.esa.int/gaia}), processed by the Gaia Data Processing and Analysis Consortium (DPAC), (\url{https://www.cosmos.esa.int/web/gaia/dpac/consortium}). 

\bibliography{cemp-no}{}
\bibliographystyle{aasjournal}

\end{document}